\documentclass[a4paper,fleqn,usenatbib]{mnras}

\usepackage[T1]{fontenc}
\usepackage{ae,aecompl}

\usepackage{graphicx}
\usepackage{color}
\usepackage{multirow}
\usepackage{amssymb,amsmath}

\def\aap{A\&A}
\def\apj{ApJ}

\def\apjl{ApJL}
\def\aj{AJ}

\def\pasa{PASA}
\def\mnras{MNRAS}
\def\nat{Nature}
\def\na{New Astron.}

\def\apjs{ApJS}

\def\ps{$\rm km \,s^{-1}\,kpc^{-1}$}

\def\arcdeg{$^\circ$}
\def\kms{\,km\,s$^{-1}$}
\def\H2{$\rm H_2$}

\title[Spiral structures in perturbed galaxies]{Gas and stellar spiral structures in tidally perturbed \\ disc galaxies}

\author[A. R. Pettitt, E. J. Tasker, J. W. Wadsley]{Alex R. Pettitt$^{1}$\thanks{E-mail:
alex@astro1.sci.hokudai.ac.jp}, Elizabeth J. Tasker$^{1}$, James W. Wadsley$^{2}$\\
$^{1}$Department of Physics, Faculty of Science, Hokkaido University, Sapporo 060-0810, Japan\\
$^{2}$Department of Physics and Astronomy, McMaster University, Hamilton, Ontario, L8S 4M1, Canada\\
}

\date{\\ \today, accepted by \mnras{} 9 March 2015}
\pubyear{2016}

\begin{document}
\label{firstpage}
\pagerange{\pageref{firstpage}--\pageref{lastpage}} \pubyear{2016}
\maketitle

\begin{abstract}

Tidal interactions between disc galaxies and low mass companions are an established method for generating galactic spiral features. In this work we present a study of the structure and dynamics of spiral arms driven in interactions between disc galaxies and perturbing companions in 3-D $N$-body/smoothed hydrodynamical numerical simulations. Our specific aims are to characterize any differences between structures formed in the gas and stars from a purely hydrodynamical and gravitational perspective, and to find a limiting case for spiral structure generation.
Through analysis of a number of different interacting cases, we find that there is very little difference between arm morphology, pitch angles and pattern speeds between the two media. The main differences are a minor offset between gas and stellar arms, clear spurring features in gaseous arms, and different radial migration of material in the stronger interacting cases. We investigate the minimum mass of a companion required to drive spiral structure in a galactic disc, finding the limiting spiral generation cases with companion masses of the order $1\times 10^{9}M_\odot$, equivalent to only 4\% of the stellar disc mass, or 0.5\% of the total galactic mass of a Milky Way analogue.
\end{abstract}


\section{Introduction}
The mechanism driving and maintaining the spiral arm structure in disc galaxies is not well understood. The problem stems from the observation that material within a galaxy orbits the centre with a frequency that decreases with radius. Any observed spiral patten will therefore wind up in such a disc (the ``winding problem").
The prevailing theory that attempts to explain the spirality is that the pattern is wave-like rather than material, with stars and interstellar medium (ISM) gas flowing in and out of the arms \citep{1964ApJ...140..646L,1973PASAu...2..174K}. While these original quasi-stationary density wave theories had the problem of spiral decay, later work on global spiral mode theories allowed for more steady spiral structure \citep{1989ApJ...338...78B,1996ssgd.book.....B}. The pattern speed of such spiral arms (the speed the arm is seen to rotate irrespective of the rotation curve) rotates with some fixed frequency; $\Omega_p={\rm constant}$. While compelling in theory, these density waves have proven difficult to definitely affirm as the explanation of spirals in all external galaxies or be reproducible in numerical simulations. 

In observations of external galaxies certain tracers should appear offset to others in and around the spiral arms (e.g. emission in CO and H$\alpha$). This is due to the shocking of the gas as it approaches the bottom of the stellar potential well, be it from up-stream or down-stream \citep{1968IAUS...29..453F,1969ApJ...158..123R}. Offsets between different galactic components have been observed in some external galaxies, but not all \citep{2009ApJ...697.1870E,2013ApJ...763...94L,2013ApJ...779...42S}. Numerical simulations have observed such offsets between stars and gas only in instances when the spiral pattern is driven by an underlying potential rather than a live stellar disc, in so-called ``dynamic spirals" \citep{2008MNRAS.385.1893D,2011ApJ...735....1W,2012MNRAS.426..167G,2015PASJ...67L...4B}.

Spiral density wave (SDW) theory also predicts pattern speeds that are constant throughout the radius of the disc. Once again this is seen in some galaxies, but more recently galaxies are seen to have radially decreasing pattern speeds \citep{2008ApJ...688..224M,2012ApJ...752...52S}. Dynamic spiral arms are however material in nature with pattern speeds that are the same as the rotation frequency of the material in the disc (e.g. \citealt{2013ApJ...763...46B,2013A&A...553A..77G,2015MNRAS.449.3911P}). 
As they exhibit material-like rotation, these arms will also wind up over the order of a galactic rotation yet are recurrent as well as transient, with new arms forming continually. Whether such a system has three arms, five arms or is near flocculent is down to the mass ratios of the various galactic components (in general, low disc to halo mass ratio systems will form more flocculent-like structures). See \citet{2011MNRAS.410.1637S} and \citet{2014PASA...31...35D} for a more in depth review of the current standing of spiral generation.

Grand design, unbarred two-armed spirals, however, present more of an issue. While a large fraction of spirals appear two armed ($\approx 50$\%), the degree of the strength and dominance of this spirality is widely variable \citep{1987ApJ...314....3E,1995ApJ...447...82R}.
 Long lasting two-armed spirals have been thus far elusive in isolated $N$-body simulations \citep{1994A&A...290..785D,1996ApJ...457..125Z,2011MNRAS.410.1637S}. While two-armed spirals can be generated, the discs tend to alternate between a two and three-armed structure or exhibit two-armed structure for only a short time-frame. There are believed to be two main alternative causes for two-armed spiral generation. The first is the rotation of an inner bar which is the likely cause for spiral arms galaxies such as NGC1300 and NGC1365 (such bars are easily generated in simulations; \citealt{2010ApJ...720L..72S,2012MNRAS.426..167G,2013MNRAS.429.1949A}). These arms tend to be fairly tightly wound and circularize at the Outer Lindblad Resonance; OLR (e.g. \citealt{2008MNRAS.388.1803R}). The nature of the arms in barred galaxies is fairly complex, with observed offsets between bar ends and spiral arms, with the two appearing dynamically decoupled in simulations \citep{2015MNRAS.454.2954B}. 
Another possible mechanism for two-armed spiral generation in disc galaxies is the tumbling or relaxation of their dark matter haloes \citep{2012ARep...56...16K,2015arXiv150701643H}. The non-axisymmetric distortions of the halo can induce two-armed spirals in embedded rotating discs, though this is an idea still in its infancy with many unknown variables and is more difficult to prove that the other mechanisms. 

The other main mechanism for generating two-armed spirals is the interaction with a companion galaxy, where the tidal force of a passing companion induces a bridge-tail structure in the host that evolves into a symmetric two-armed spiral \citep{1972ApJ...178..623T,1991A&A...252..571D}.
Interactions such as these are believed to be responsible for some of the most well known two-armed spiral galaxies. The poster-child of grand design spiral structure, M51, is clearly interacting with the smaller NGC\,5195 and the system has been reproduced with simulations in numerous previous studies (e.g. \citealt{2000MNRAS.319..377S,2010MNRAS.403..625D}). The structure of our nearest neighbour M31 is difficult to discern due to the large inclination on the night sky, however \citet{2014ApJ...788L..38D} find that a penetrating orbit of a small companion from above the galaxy can induce the somewhat irregular spiral morphology seen in observations. The grand-design spiral M81 is part of a more complex interacting system, with the tidal interactions of at least two other nearby bodies believed to be driving the spiral structure, making reproduction with simulations a difficult endeavour \citep{1993A&A...272..153T,1999IAUS..186...81Y}. The Milky Way itself has several nearby galactic-sized objects, most notably the Large and Small Magellanic Clouds. While the satellites of the Milky Way are not believed to be the sole drivers of the observed spiral structure, they can explain some of its morphological features \citep{2009MNRAS.399L.118C,2011Natur.477..301P}. There are also many less well known galaxies that are believed to have structures driven by companions, be it a minor companion fly-bys such as NGC\,2535 or NGC\,6907, or a more grandiose encounter situation such as NGC\,6872 or Arp\,273.

Previous theoretical and numerical work on tidal encounters in galactic systems has primarily focused on $N$-body stellar simulations, beginning with the seminal work of \citet{1972ApJ...178..623T}. 
Tidal forces tend to scale as $F_{\rm tide} \propto M_p/d^3$, where $M_p$ and $d$ are the mass and pericentric distance to the perturbing companion. This means some degeneracy exists between the $M_p$ and $d$, though methods have been suggested that can break this degeneracy \citep{2009MNRAS.399L.118C}. The velocity vector of the companion plays a more indirect role, with tidal features being strongest when the velocity of the companion is comparable to the rotation speed of the host galaxy near closest approach \citep{2010ApJ...725..353D}.
Tidal encounters can be seen to drive many different morphological features, depending on the properties of the interaction. 
This includes driving \citep{1998ApJ...499..149M,2014ApJ...790L..33L} and hindering bar formation \citep{2008ApJ...687L..13R}, creating ringed and spoked features  \citep{2012MNRAS.425.2255F,2015ApJ...805...32W}, and generating grand-design two-armed spirals \citep{1992AJ....103.1089B,1998ASPC..136..260D,2008ApJ...683...94O,2011MNRAS.414.2498S}. 
Little work has been done on a cosmological perspective on the tidal driving of spiral structures. Simulations of a halo filled with dark satellites by \citet{2008ASPC..396..321D}, seeded by dark matter simulation statistics of \citet{2004MNRAS.355..819G}, find that spirals can be easily generated in a subhalo passage, though these are short-lived.

There exists several studies that focus of the driving of spirals in galactic discs in tidal encounters. In the study of \citet{1992AJ....103.1089B} the authors find a lower limit on galaxy to companion mass ratio of 0.01 for driving spiral structure, though their calculations are limited to a static halo and confined to 2D. Similar lower mass limits were used in the simulation surveys of \citet{2011ApJ...743...35C}, and \citet{1991A&A...244...52E}, though in these cases no detailed morphological study was shown for the varying interactions.
\citet{1996ApJ...471..115B} include gas in their simulations, but no detailed morphological study, and are fairly low in resolution. \citet{2011MNRAS.414.2498S} also include gas, but limit their study to only three calculations, and find little to no spiral structure induced in a 0.01 companion to galaxy mass ration encounter.
The studies of \citet{2008ApJ...683...94O} and \citet{2015ApJ...807...73O} offer the closest analogy to the work presented here, looking into the arm structure and dynamics in several simulations, however, they include no gas in their calculations.

In the studies mentioned above we find two questions to be unanswered. The first is how do the gas and stellar morphologies differ in different interactions? Will their morphology be the same in interactions with different masses, orbital inclinations, or will they evolve similar structures regardless of the specifics of the interaction? We specifically aim to look at serval key quantities of the structures in the gas and stellar components; the pattern speed, arm number, pitch angle and radial migration to ascertain whether there is any differences between the two that may in turn be of use to detect such interactions in external galaxies. We also study any offsets between the two media, any such instance is of interest due to the appearance of such features in density wave driven arms and dearth of in the dynamic spiral case. 
The second key point is finding how small the companion can be and still trigger a spiral in the disc. This has serious observational consequences, and is important for placing limits on companions that could be responsible for unbarred two-armed spirals (such as NGC\,1566, NGC\,2535 and NGC\,2857).

This paper is organized as follows. Sec. \ref{sec:numerics} describes the setup and technical details of the calculations. We briefly first discuss the initial isolated disc in Sec. \ref{Res1}, before the introduction of a companion. We then discuss our fiducial simulation in detail in Sec. \ref{Res2}. The results of the parameter study are presented in Sec. \ref{Res3} and we conclude in Sec. \ref{conclusions}. The appendix includes a brief analysis of supplementary models that are permutations of our fiducial model (e.g. different resolutions and gas temperature).

\section{Numerical Simulations}
\label{sec:numerics}
\begin{figure}
\includegraphics[trim = 7mm 10mm 0mm 5mm,width=90mm]{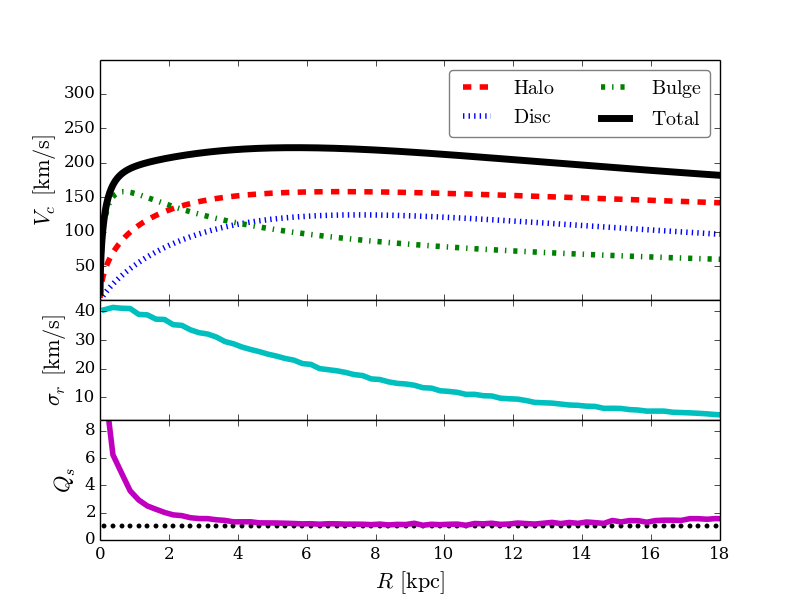}
 \caption{The rotation curve (top), initial radial velocity dispersion (middle) and initial Toomre-$Q$ parameter in the stars in the simulations presented in this work. Parameters governing each component are given in Table\;\ref{ICtable}.}
\label{RC}
\end{figure}

The simulations presented here focus on the case of a small companion interacting with a host disc galaxy. The galaxy is composed of an $N$-body stellar disc, spherical inner bulge and outer halo. ISM gas is also included in the disc, with a stellar to gas mass ratio of 10:1. Initial conditions for the isolated galaxy are generated using the \textsc{nemo} stellar dynamics toolbox \citep{1995ASPC...77..398T}, specifically the \textsc{magalie} initial conditions generator \citep{2001NewA....6...27B}, itself based on the method of \citet{1993ApJS...86..389H}. The profiles describing the various components are an exponential stellar disc, a truncated isothermal dark matter halo and a Hernquist-type bulge. The various masses and scale lengths of each of the components are listed in Table\;\ref{ICtable}.
 Fig.\,\ref{RC} shows various properties of the initial conditions as a function of galactic radius including the individual rotation curve components (top) the radial velocity dispersion (middle) and Toomre-$Q$ parameter in the stars (bottom), given by the equation;
 \begin{equation}
Q_s = \frac{\kappa \sigma_R}{3.36 G \Sigma_0}
\end{equation}

where $\kappa$ is the epicycle frequency, $\sigma_R$ the radial velocity dispersion and $\Sigma_0$ the disc surface density. Values of $Q_s<1$ imply the disc is gravitationally unstable \citep{1964ApJ...139.1217T}. The rotation curve is tailored to represent a general disc galaxy, with a velocity amplitude of approximately 200\kms{}, peaking at approximately 220\kms{} at $R=3{\rm kpc}$. The orbital period at two disc radial scale lengths ($2a_{\rm disc}=7{\rm kpc}$) is approximately 190 and 540\,Myr at the disc edge ($R=20$kpc).

The initial galaxy is designed to be near-flocculent to make any arm the result of the interaction rather than the disc's own instabilities. This requires a rotation curve that is halo dominated, in accordance with the predicted swing amplification mode (see \citealt{1981seng.proc..111T}, \citealt{2011ApJ...730..109F}, \citealt{2015MNRAS.449.3911P}). This, in addition to the relatively heavy inner bulge, also makes the disc stable to bar formation due to the establishment of a stabilizing Toomre $Q$-barrier in the inner disc. Bars are undesirable in this study as they would make it difficult to discern whether arms are driven by the interaction or by the bar rotation \citep{1987gady.book.....B,1995gaco.book.....C}. This galaxy model is very similar to the Bd model in \citet{2015MNRAS.449.3911P} except with a live halo. A live halo is necessary to correctly model strong interactions, as a static halo would produce an unrealistic anchor for the disc-bulge system, though this does introduce additional dynamical friction.
Our fiducial resolution is $2\times 10^4$ bulge, $3\times 10^5$ disc, $3\times 10^5$ gas and $3\times 10^5$ halo particles, though we also perform calculations with 1 million disc particles as a resolution check. Our fiducial resolution is lower than is advised by some literature studies to resolve bar \citep{2009ApJ...697..293D} or spiral \citep{2011ApJ...730..109F} features driven by self-gravity in the stellar disc. However, we stress that are not wishing to follow such structures, and instead wish for a featureless disc prior to the companion interaction.
The perturbing companion is normally modelled as a single heavy dark matter particle, though also we run a single computation using a resolved Plummer sphere to represent the companion.

\begin{table}
\centering
 \begin{tabular}{@{}l | c l}
  \hline
  Param & Value & Desc.\\
  \hline
  \hline
  $n_{\rm halo}$ & $3\times 10^5$ & Halo particle number \\
  $n_{\rm gas}$ & $3\times 10^5$ & Gas particle number\\
  $n_{\rm disc}$ & $3\times 10^5$ & Disc particle number\\
  $n_{\rm bulge}$ & $2\times 10^4$ & Bulge particle number\\
  $n_{\rm pert}$ & 1 or $2\times 10^4$ & Perturber particle number \\
  $\epsilon_{\rm soft}$ & 50pc & Grav. softening length\\
  \hline
  $M_{\rm disc}$ &  $3.0 \times 10^{10}$ $M_\odot$ & Mass of stellar exponential disc \\
  $a_{\rm disc}$ & 3.5\,kpc & Scale length of stellar exponential disc \\
  $r_{t,\rm disc}$ & 20\,kpc  & Truncation length of disc \\
  $M_{\rm halo}$ & $17 \times 10^{10}$ $M_\odot$ & Mass of isothermal halo\\
  $a_{\rm halo}$ & 7\,kpc & Scale length of isothermal halo \\
  $r_{t,\rm halo}$ & 60\,kpc  & Truncation length of halo \\
  $M_{\rm bulge}$ & $1.6 \times 10^{10}$ $M_\odot$ & Mass of Hernquist bulge\\
  $a_{\rm bulge}$ & 0.7\,kpc & Scale length of Hernquist bulge\\
  $M_{\rm gas}$ & 0.1$M_{\rm disc}$& Mass of gas \\
  $M_{ p,0}$ &$2 \times 10^{10}$ $M_\odot$& Fiducial mass of perturber \\
  \hline

 \end{tabular}
 \caption{Fixed values in our simulations, including resolutions and parameters governing the rotation curve of the primary galaxy.}
 \label{ICtable}
\end{table}

The orbit is set to be parabolic for our default calculation, with a closest approach of approximately 20kpc and an initial velocity magnitude of 50\kms. The companion is initially 140kpc away from the host galaxy, to ensure it is well outside of the majority of the halo galactic mass distribution upon inclusion in the system. We focus our efforts on grazing minor interactions, where the companion is at least an order of magnitude less massive than the host, being physically analogous to a small dwarf galaxy or dark matter subhalo. This initial configuration was chosen by performing a series of lower-resolution simulations to find a configuration that produced a strong two-armed perturbation while keeping a grazing closest approach and an initial distance well outside the majority of the halo mass distribution. We investigate the effect of changing the mass, velocity, closest approach and orbital path on the morphology of the host galaxy. Note that due to the effect of dynamical drag as the companion passes through the halo, the orbits may become strongly bound regardless of the seeded orbit. The separate calculations are listed in Table\;\ref{RunParams}, which includes several different companion masses, two resolution tests and four different orbital inclinations. 
The Extended model has an extended gas disc and larger-scale length (a factor of 2 increase for both compared to the fiducial run) to mimic observations of spiral galaxies with extended gas discs. The mass is the same as the normal runs, so the effective surface density is lower and the path of the companion now directly travels through the gas disc for a nearly half a galactic rotation.
Also included are two calculations where two-armed spirals are generated without a companion, including SDW and dynamic, transient and recurrent (DTR) spirals.

Simulations were performed using the $N$-body+smoothed particle hydrodynamics (SPH) code \textsc{gasoline} \citep{2004NewA....9..137W}. Gravity is solved using a binary tree, and the system integrated using a kick-drift-kick leapfrog. Self gravity is active for all components, using a fixed gravitational softening of 50pc. The gas is isothermal with a temperature of 10000K, in effect simulating the warm ISM. An additional calculation with 1000K gas (DefaultCld) is included, but to avoid large-scale collapse on the scale of the gravitational softening length we use a surface density half that of the fiducial calculation. Calculations with a 200K gas were initially included, but the disc experienced wide scale collapse rapidly even before the perturber passage. With additional physics such as star/sink formation and supernova feedback the calculation could continue, but for the simulation survey conducted here we prefer to omit these additional physical processes. These will instead be used in a future study with a much smaller number of simulations and higher resolution.
 The different permutations of the Default calculation (DefaultX3, Resolved, Extended, DefaultCld) are primarily discussed in the appendix and only mentioned briefly in the main text. We found the different resolution tests do not change the driven spiral features. The gas distribution and temperature do have some effect on arm features, reasons for which are discussed in the appendix.

\begin{table*}
\centering
 \begin{tabular}{@{}l c c c c c l }
  \hline
  Model & $M_p\, {\rm[ 2\times10^{10} M_\odot]}$ & $V_\circ$ [\kms] & $\theta\, {\rm[deg]}$ & $\psi\, {\rm[deg]}$ &  $d {\rm[kpc]}$  & Notes\\
  \hline
  \hline
  Default & 1 & 50 & 0 &0 & 20 &  Our fiducial calculation\\
  DefaultX3 & 1 & 50 & 0 &0 & 20 &  As above with triple resolution in galaxy\\
  Resolved & 1 & 50 & 0 &0 & 20 &  As fiducial run but with resolved companion\\
  Extended & 1 & 50 & 0 &0 & 20 &  As fiducial run but with an extended gas disc\\
  DefaultCld & 1 & 50 & 0 &0 & 20 &  As fiducial run but with 1000K gas disc and $0.5M_{\rm gas}$\\
  \hline
  Heavy1 & 2 & 50 & 0 &0 &20  & Heavier companion than the fiducial run\\  
  Light1 & 0.5 & 50 & 0 &0 &20 & Companion 50\% mass of Default\\
  Light2 & 0.25 & 50 & 0 &0 &20 & Companion 50\% mass of Light1\\
  Light3 & 0.125 & 50 & 0 &0 &20 &  Companion 50\% mass of Light2\\
  Light4 & 0.0625 & 50 & 0 &0 &20 & Companion 50\% mass of Light3 \\
  Light4d1 & 0.0625 & 50 & 0 &0 & 16 & As Light4 but reduced closest approach\\
  Light4d2 & 0.0625 & 50 & 0 &0 & 12 &  As Light4d1 but reduced closest approach\\
  Light4d3 & 0.0625 & 50 & 0 &0 & 8 & As Light4d2 but reduced closest approach \\
  \hline
  Orbit45 & 1 & 50 & +45 &0 &20 & Perturber orbit is $+45^\circ$ out of plane\\
  Orbit90 & 1 & 50 & +90  &0 &20&  Perturber orbit is $+90^\circ$ out of plane (follows $x=0$)\\  
  Orbit135 & 1 & 50 & +135 &0 &20 & Perturber orbit is $+135^\circ$ out of plane\\
  Above & 1 & 50 & 0 &+90 &20 &  Perturber initially above North Galactic Pole\\
  \hline
  Slow1 & 1 & 40 & 0 &0 &20 & Perturber velocity has additional $-10$\kms{} initially\\
  Fast1 & 1 & 60 & 0 &0 &20 & Perturber velocity has additional $+10$\kms{} initially\\  
  \hline
 SDW &- & - & - & - & - & Gas disc with analytic disc and spiral potentials \\  
 DTR & - & - & - & - & - & Isolated galaxy with $\times 2 M_d$\\  
  \hline
  \hline
 \end{tabular}
 \caption{Description of the calculations presented in this work. The latter two calculations have no perturbing companion, and are included to illustrate SDW and DTR two-armed structures.
 The parameters describing the companion are the closest approach of the orbit, $d$, the initial velocity magnitude, $V_\circ$, and mass of the perturber, $M_p$. The angle $\theta$ measures the rotation of the initial velocity vector about the $y$-axis (i.e., $\theta=180^\circ$ is a retrograde orbit) and $\psi$ is the rotation about the $x$-axis (i.e. when $\psi=90^\circ$ the companion originates from the North Galactic Pole).}
 \label{RunParams}
\end{table*}

\section{Results and discussion}
\label{sec:results}

\subsection{Initial isolated disc}
\label{Res1}

\begin{figure}
\centering
\resizebox{1.0\hsize}{!}{\includegraphics[trim = 0mm 0mm 0mm 0mm]{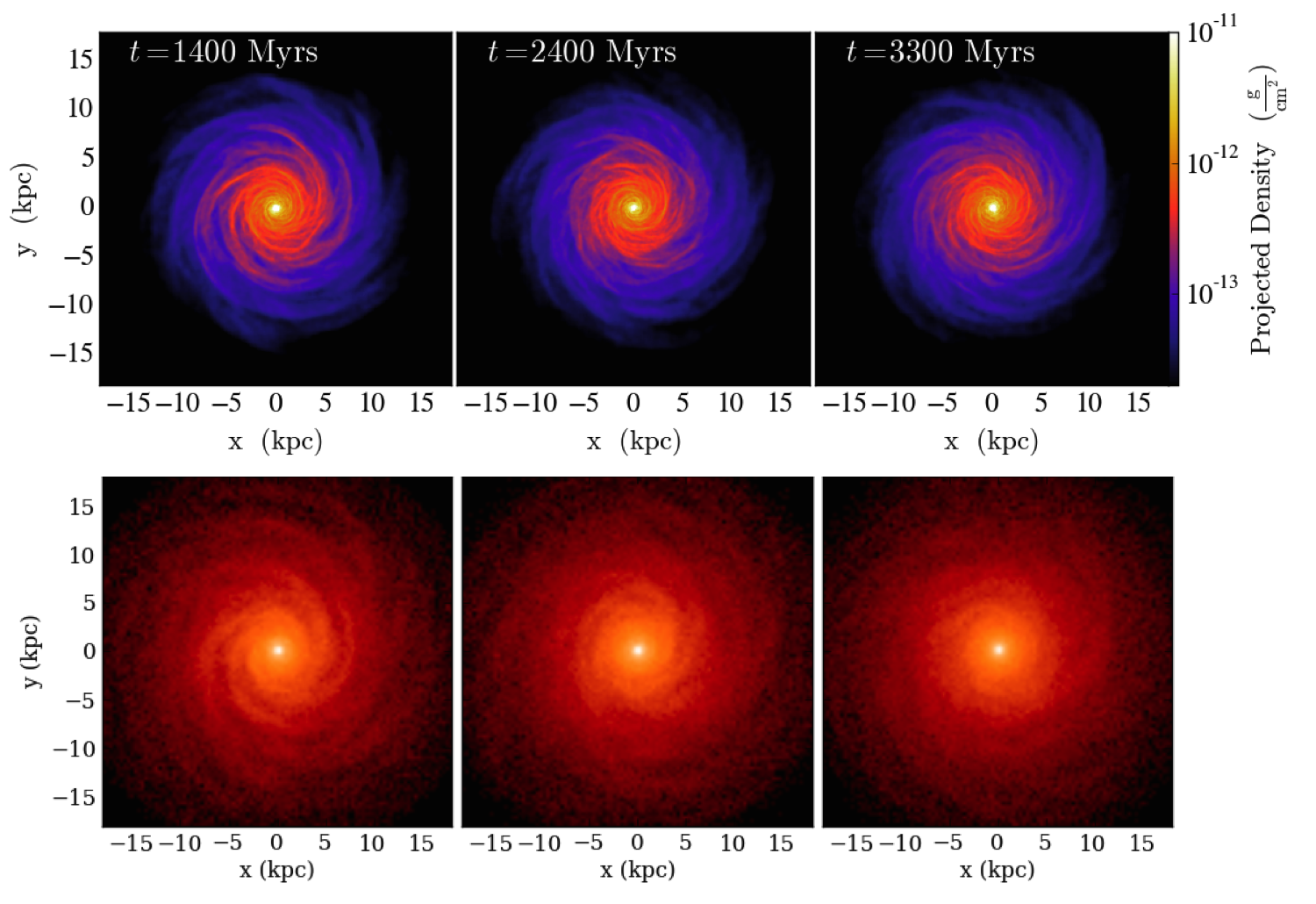}}
  \caption{Density projection of the gas (top) and position of star particles (bottom, no bulge stars plotted) of the initial isolated galaxy at three separate times, showing general global stability.}
  \label{IsoDisc}
\end{figure}

The evolution of the isolated disc is shown in Fig.\,\ref{IsoDisc} in the gas (upper) and stellar (lower) components, where times are given from the time of initialization. There is some structure in the disc at early times ($t\approx 1$\,Gyr) but this dies away with time. The companion does not reach the stellar disc until after $t\approx 2$\,Gyr, and while there is still some structure in the disc at this stage, the structure is near flocculent and well smoothed out. Fourier mode analysis of the disc in this time frame shows the isolated galaxy has a dominant arm mode of approximately $m=5$, but this is comparable to the noise of the other modes and is highly time-dependant.

We allow the initial galaxy to evolve for 1\,Gyr before the inclusion of the perturber into the system.
The actual interaction does not occur for another Gyr due to the companion being placed over 100kpc away. Over this time the Toomre Q parameter in the stars rises from an initial value of 1.2 to approximately 1.8 in the mid disc over the snapshots shown in Fig.\,\ref{IsoDisc}.

\subsection{Fiducial simulation}
\label{Res2}
\begin{figure*}
\centering
\resizebox{0.9\hsize}{!}{\includegraphics[trim = 0mm 0mm 0mm 0mm]{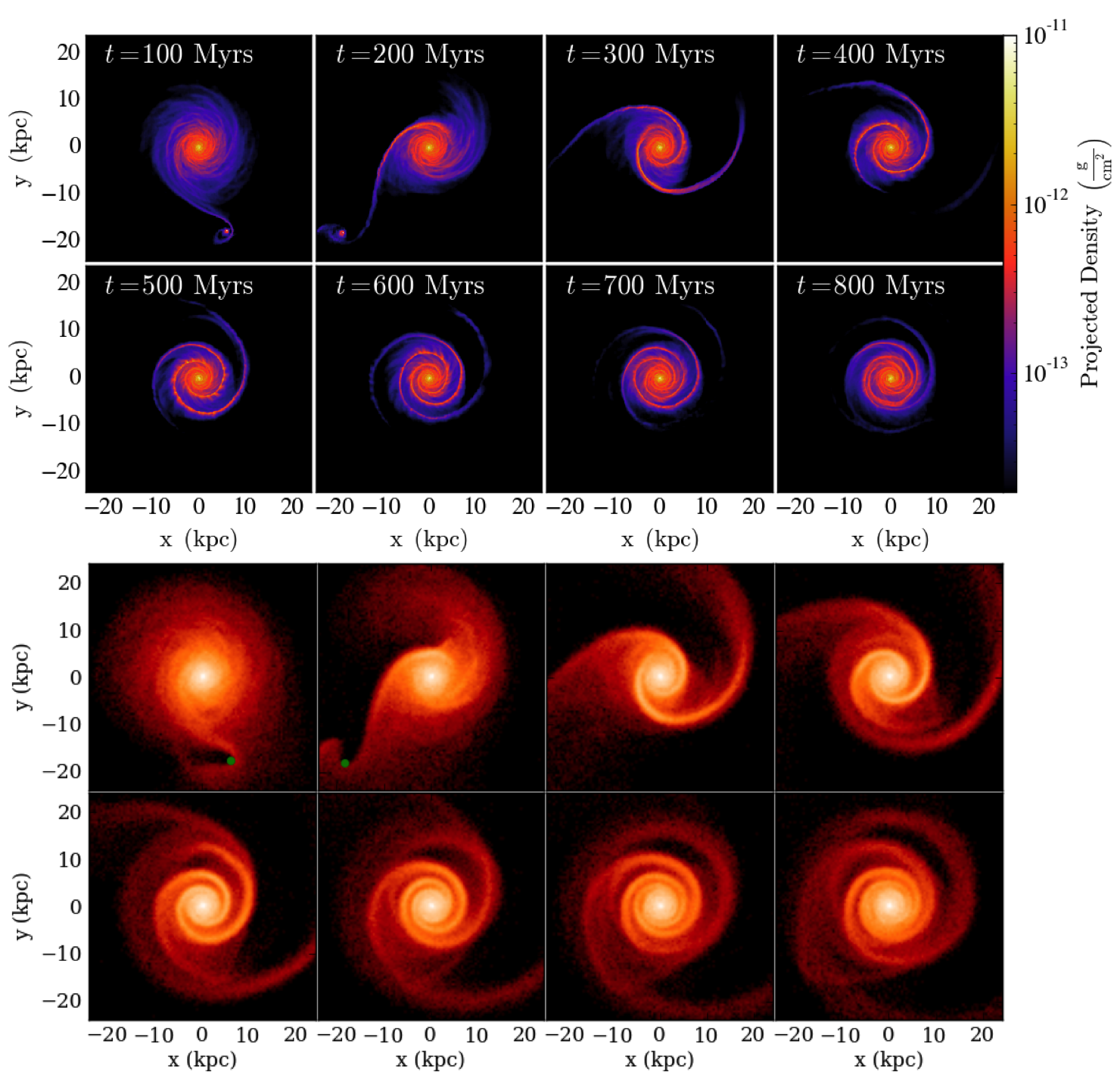}}
  \caption{Time lapse of the fiducial simulation during and after interaction with the perturber over 700\,Myr. A density render of gas is shown in the top panel, and the positions of star particles in the lower. The galaxy is rotating clockwise and the position of the small companion is indicated by the green dot. The (0,0) location is the centre of mass of the bulge component.}
  \label{PertLapse}
\end{figure*}

In Fig.\,\ref{PertLapse} we show a time-lapse of our fiducial simulation. The top panels show the gas surface density and the lower panels the position of the disc star particles spanning a time of 700\,Myr. The companion is indicated by the green point in the first two panels of the stellar distribution, and originates from the top of the page (it then moves out of frame in following panels). The galaxy is initially positioned at (0,0,0) but experiences a drift due to unanchored halo and attraction of the companion. The images in Fig.\,\ref{PertLapse} have been re-centred to the bulge centre of mass for clarity. A clear bridge-tail system is driven soon after the encounter which evolves into a two-armed spiral pattern after approximately half a rotation (150\,Myr). The bridge experiences a bifurcation before the transformation into more regular spiral structures (seen more clearly in the stellar material). The gas in general traces a much finer armed structure, while the stellar distribution is smoother and has more inter-arm material. The arm features persist for two galactic rotations (600\,Myr) before they become distorted from an ideal log spiral structure. After the encounter the perturbing companion stripped the host galaxy of a small amount of gas and stellar material (4\% and 3\% respectively) while the dark matter material is less affected.

\begin{figure*}
\centering
\resizebox{0.80\hsize}{!}{\includegraphics[trim = 0mm 0mm 0mm 0mm]{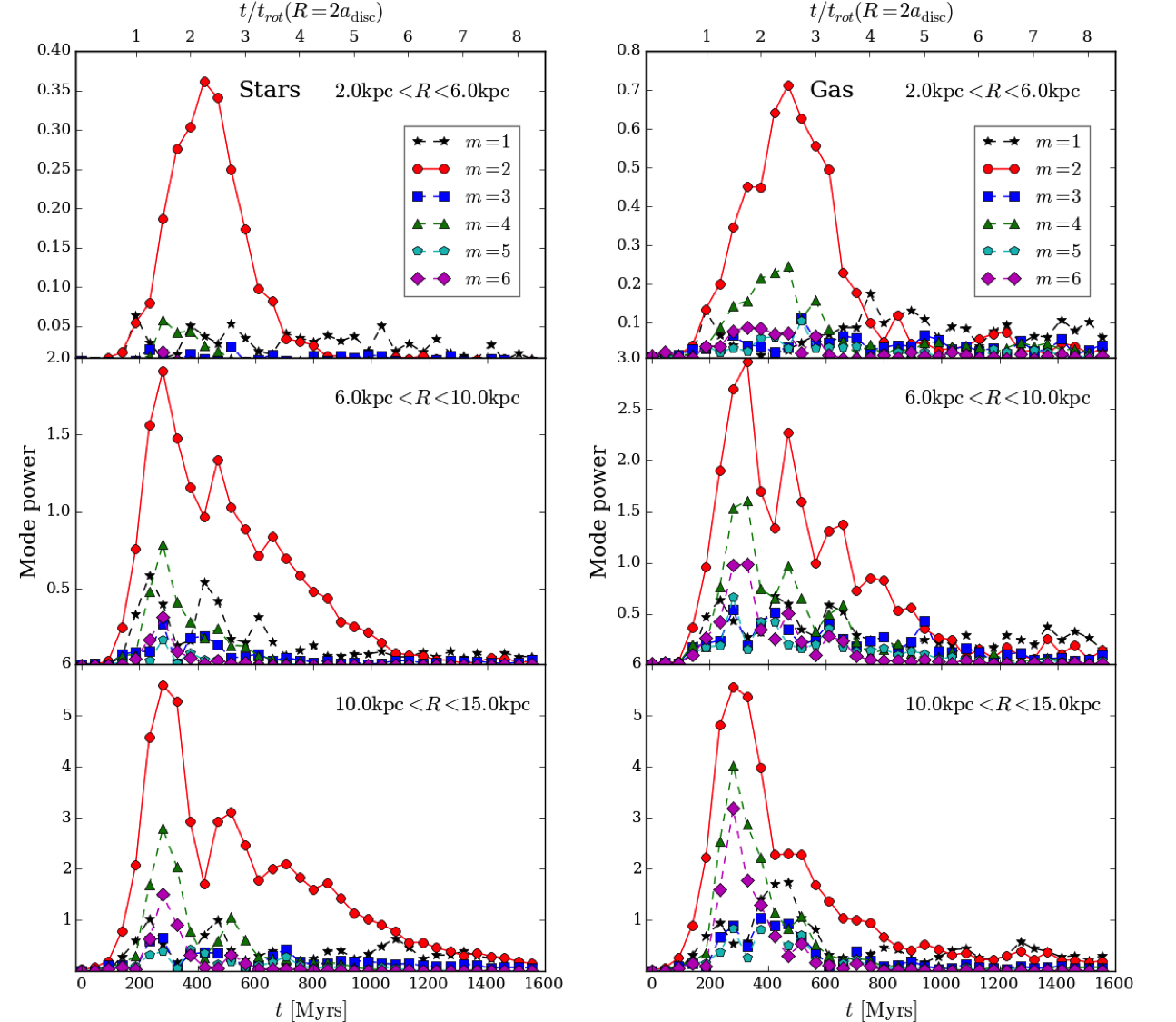}}
  \caption{Evolution of the power of various Fourier modes in our fiducial simulation, shown in Fig.\,\ref{PertLapse}. The stellar component is shown on left and the gas on the right. Each column shows a different radial region; the inner disc (2kpc$<R<$6kpc), the mid-disc (6kpc$<R<$10kpc) and the outer disc (10kpc$<R<$15kpc).}
  \label{ArmModes}
\end{figure*}

To further quantify the features generated by the tidal interaction, we perform a Fourier analysis of the stellar and gaseous material. This enables the calculation of the dominant arm number, which in turn can be used to infer the pitch angle (the angle an arm makes with a tangent to a circle at the same radius, $\alpha$) and pattern speed (the rotation speed of the observed spiral structure, $\Omega_p$) of the spiral arms. The details of this calculation are described in the appendix of \citet{2015MNRAS.449.3911P}.

Fig.\,\ref{ArmModes} shows the power in each Fourier arm mode ($m$) for our fiducial model as a function of time after the passage of the companion. As the response of the disc is different at each radius, we show the mode power at an inner ($2{\rm kpc}<R<6{\rm kpc}$), mid ($6{\rm kpc}<R<10{\rm kpc}$), and outer ($10{\rm kpc}<R<15{\rm kpc}$) disc region. The stellar material is shown in the left-hand column, and the gas in the right. The panels show the time from just after periastron passage of the companion, to when the disc appears to have reverted to its a structure similar to the original morphology, at approximately 1.6Gyr. There is still some remnant power in the $m=2$ and $m=1$ modes after this time, but it is very weak, highly irregular in structure, and confined to the outer disc. There is a clear increase in the power of the $m=2$ mode throughout the disc after the interaction, which appears to dominate at all radii for over 1\,Gyr. This then slowly dies away to powers similar to the other arm modes. The stars have a generally clearer peak in the $m=2$ mode, while the gas has more power allocated to other modes due to its more filamentary nature. There is a small dip in the power of the $m=2$ after the primary peak, which corresponds to when the bridge bifurcates, seen clearly in the stars during $400 {\rm Myr} \le t \le 500 {\rm Myr}$ in Fig.\,\ref{PertLapse}.
There is some additional power in the other even modes (4 and 6). There are two possible reasons for this. The first is that there is genuine power in these modes in the disc. In inspection of Fig.\,\ref{PertLapse} there is some bifurcation of two to four armed structure in disc, seen clearly in the 400\,Myr timestamp, which would explain some of the additional power in the $m=4$ mode around the same time. The other reason is the square wave-like nature of the density structure with azimuth at each radii, which will boost the power in other even modes for a $m=2$ signal.

\begin{figure}
\centering
\resizebox{1.0\hsize}{!}{\includegraphics[trim = 0mm 0mm 0mm 0mm]{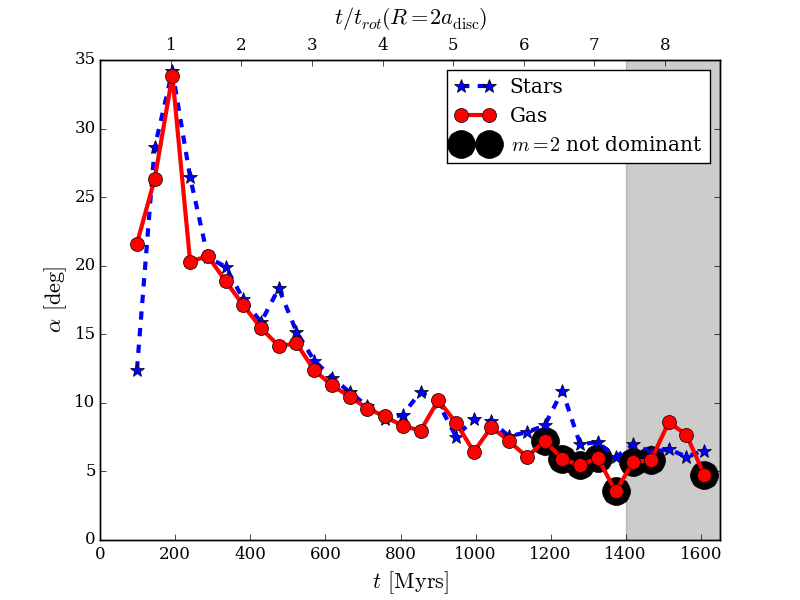}}
  \caption{The evolution of the pitch angle for our fiducial simulation for the $m=2$ mode. The blue and black lines show the fit to the stellar and gaseous arms respectively. The black points indicate times where the $m=1$ mode had the greater power. The grey region indicates the time frame where the power of the $m=2$ mode becomes comparable to the other modes (i.e. noise).}
  \label{Alpha}
\end{figure}

With the dominant modes in the disc traced, and clearly belonging to the $m=2$ family, we then fit logarithmic spiral functions to the gas and stellar arms. The resulting pitch angles of the spiral arms in the gas and stars are shown in Fig.\,\ref{Alpha} as a function of time. There is a clear decrease in pitch angle with time, indicating the arms winding up, and follows an exponential-like decay. The black points indicate the fits where the $m=2$ mode is not the dominant mode, and only occurs for the gas where the $m=1$ mode (a large one-armed feature in the outer disc) begins to take over at later times. The shaded region shows approximately where the $m=2$ power decreases to the ambient level of all the other modes, thus making a fit and pitch angle determination problematic. The maximum for both media is near 35\arcdeg{} and reaches a minimum of about 6\arcdeg{}. These span almost the entire range of values seen in external galaxies (e.g. \citealt{1998MNRAS.299..685S}). This is also the only way of producing two-armed spirals in simulations with comparatively small pitch angles, whereas isolated galaxy simulations with DTR arms tend to favour wide pitch angles of 15\arcdeg{}$\le \alpha \le$30\arcdeg{} \citep{2013A&A...553A..77G,2015MNRAS.449.3911P}.

\begin{figure}
\centering
\resizebox{1.\hsize}{!}{\includegraphics[trim = 0mm 8mm 0mm 0mm]{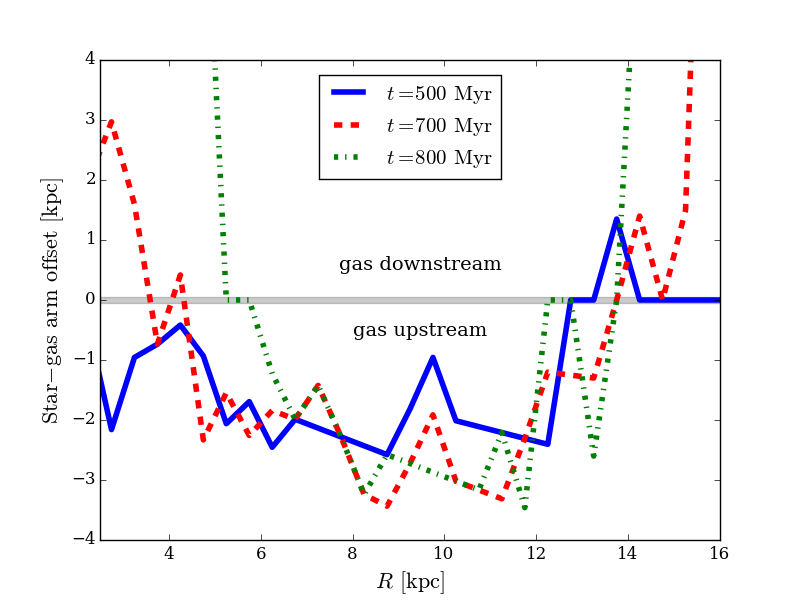}}
  \caption{The offset between the spiral arms in the gas and stars in the fiducial simulation as a function of radius at three different times after the periastron passage. A minor offset can be seen between the stars and gas at all times in the mid-disc. The large differences in the inner and outer disc are partially due to the near featureless disc centre and the lower surface density in the outer disc.}
  \label{Offs}
\end{figure}

There is some evidence and interest in whether different galactic components trace different structures. For example, in the M51 PAWS data \citep{2013ApJ...779...42S} offsets can be seen between the star formation regions, molecular gas and old stellar population. Even in our own galaxy there is some evidence that gas and stars trace entirely different arm numbers \citep{2001ApJ...556..181D}. While the dominant arm mode is clearly $m=2$ in these calculations, we can assess any offsets between components. 
In Fig. \ref{Offs} we show the offset in the spiral features traced by the stars and the gas in our fiducial simulation. No logarithmic spiral assumption is used, with instead the points reflecting the greatest density of material in each radial bin. The vertical axis is simply the azimuthal position of the gaseous arm subtracted from the stellar arm and multiplied by $\pi R$ to give the azimuthal distance offset. There is small yet noticeable offset between the gaseous and stellar arms in the mid/outer disc, with the gas leading the stars upstream in this region. This is significantly larger than resolution limits (gravitational softening is 50pc and  gas smoothing lengths being centred approximately around 80pc). 
The offset is significantly smaller (of the order of 1\arcdeg{}-2\arcdeg{}) than in simulations of density-wave driven spirals \citep{2015PASJ...67L...4B}, and offsets between different gas components in galaxies where a number of negligible to small-scale offsets ($<10^\circ$) are observed \citep{2013ApJ...763...94L}. As the spiral arms are only pseudo-density waves, shocking of the gas detailed in works such as \citet{1969ApJ...158..123R} should be much weaker and likely only partially causing the offset seen here. The remaining offset is likely due to the tidal nature of the interaction, in which the companion slightly tugs the gas out of stellar potential well upstream of the spiral existing spiral arms. Further investigation on offsets seen in tidal spirals in comparison to what is seen in observations is needed, which we leave to future work.

The speed that spiral arms rotate at is another point of debate in the community. The standard SDW theory assumes the spirals are rotating with some constant pattern speed, while numerical simulations with live discs (e.g. \citealt{2013ApJ...763...46B,2013A&A...553A..77G,2015MNRAS.449.3911P}) and an increasing amount of observations (e.g. \citealt{2006MNRAS.366L..17M,2008ApJ...688..224M,2012ApJ...752...52S}) show arms that are winding with a rotation speed indistinguishable from the material speed.

\begin{figure}
\centering
\resizebox{0.9\hsize}{!}{\includegraphics[trim = 10mm 0mm 10mm 0mm]{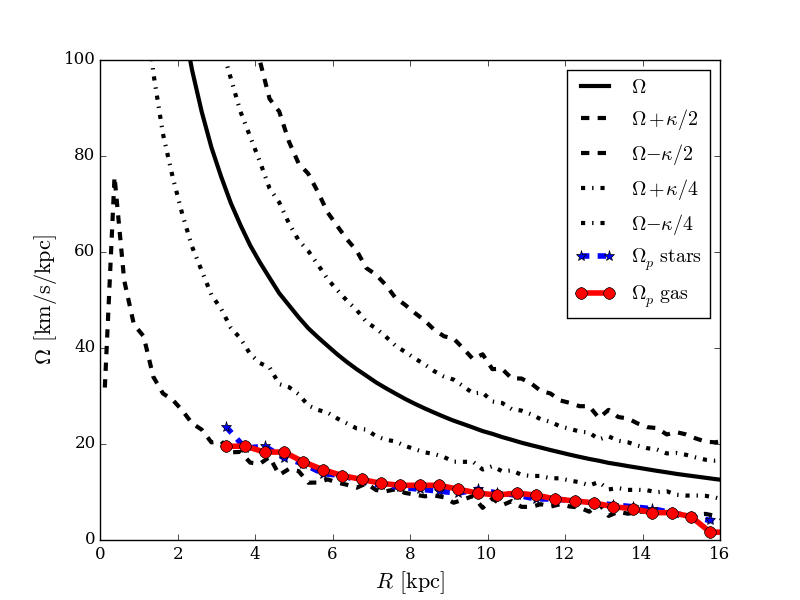}}
  \caption{The pattern speed (in the gas and stars) as a function of radius plotted against the rotation frequencies (in the stars) in the galactic disc. Pattern speeds here are calculated between 400 and 600\,Myrs.}  
  \label{PatternSpeed}
\end{figure}

\begin{figure*}
\centering
\resizebox{0.8\hsize}{!}{\includegraphics[trim = 0mm 15mm 0mm 5mm]{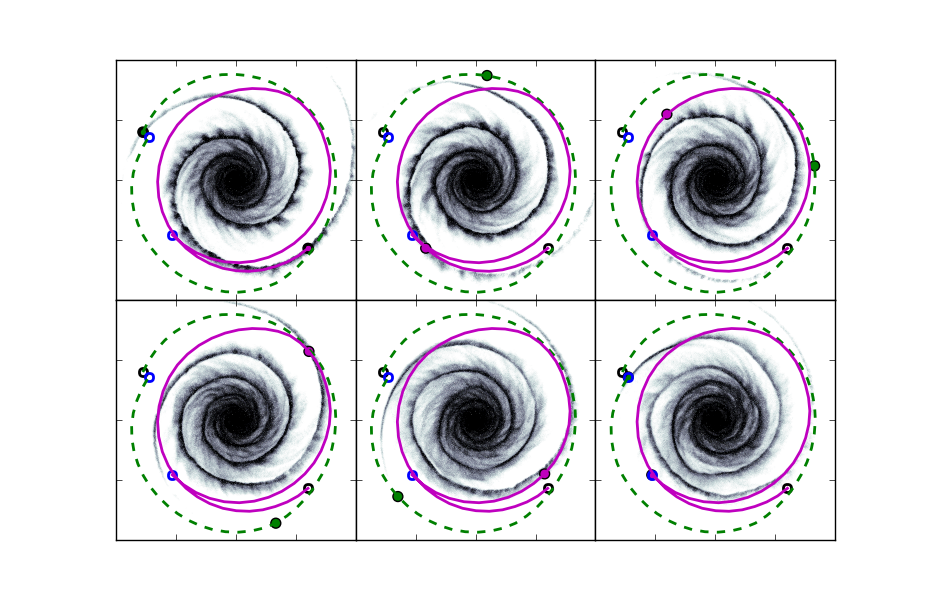}}
  \caption{Time evolution of the positions of gas particles in the host galaxy in our fiducial simulation, coloured by normalized density. The green and magenta circles show two individual gas particles as they move through the disc, and are initially co-incident with the spiral structure. The black and blue open circles are the start and end points of the paths of the two particles. These two particles can be seen to exit and re-enter the spiral arm density waves. The panels are of length 20kpc.}
  \label{TrackGas}
\end{figure*}

We measure the pattern speed, $\Omega_p$, of the spiral arms in our fiducial simulation in the stellar and gas components, the results for which are shown in Fig.\,\ref{PatternSpeed}. Pattern speeds are only shown in the range where spiral features can be clearly fit, which does not include the fairly featureless inner disc. Rotation frequencies in the disc are indicated by the black lines, including the $\Omega\pm \kappa/2$ and $\Omega\pm \kappa/4$ resonances (where $\kappa$ is the epicycle frequency and $\Omega$ the rotation frequency of the galactic material). The pattern speed is clearly not constant, as would be expected for winding arms, but it is also not exactly material ($\Omega_p \ne \Omega(R)$). The arms are wave-like, with material flowing in and out of the spiral potential well, but also experience winding due to a non-constant $\Omega_p$. This is highlighted in Fig.\,\ref{TrackGas}, where we show the evolution of the two armed spiral in the gas, and the locations of two individual gas particles (green and magenta points). The paths of the particles are traced by the solid lines as the disc evolves for  approximately a full rotation at $R=2a_{\rm disc}$.The particles are selected to be within the spiral arms initially, and can be seen to flow out of the arms, through the inter arm region and then back into another arm (starting and finishing in the black and blue open circles respectively). 

The pattern speed in Fig.\,\ref{PatternSpeed} clearly traces the $\Omega- \kappa/2$ frequencies, so there is no resulting inner or outer OLR, or corrotation radius. The gas is always moving faster than the spiral arms, and flows into the perturbation from behind. When the companion passes the point of closest approach it is moving with a velocity of 270\kms relative to that of the main galaxy. As closest approach is 20kpc, this results in a circular frequency of 13.5\ps{}, which is slightly higher than the rotation frequency in the disc at this radius ($\approx 11$\ps{}). As such, the frequency of the perturber does not need to be an exact match to any of the frequencies of the disc, be it $\Omega$ or $\Omega- \kappa/2$, to successfully drive a spiral pattern in the disc that persists for Gyr time-scales.
The time dependence of the pattern speed will be discussed in further detail in Sec. \ref{Res3}.
 
It is possible the slight offset seen in Fig.\,\ref{Offs} and the non-material yet non-constant pattern speed in Fig.\,\ref{PatternSpeed} is an indication of the middle-ground nature of these spirals. They are not quite standing density waves (with $\Omega_p={\rm const.}$ and clear gas-star offset) and not quite material arms (with $\Omega_p=\Omega(R)$ and coincident gas-star arms).

\subsubsection{Spurring features}
\label{SecSpur}

\begin{figure}
\centering
\resizebox{.45\hsize}{!}{\includegraphics[trim = 30mm -5mm 10mm 5mm]{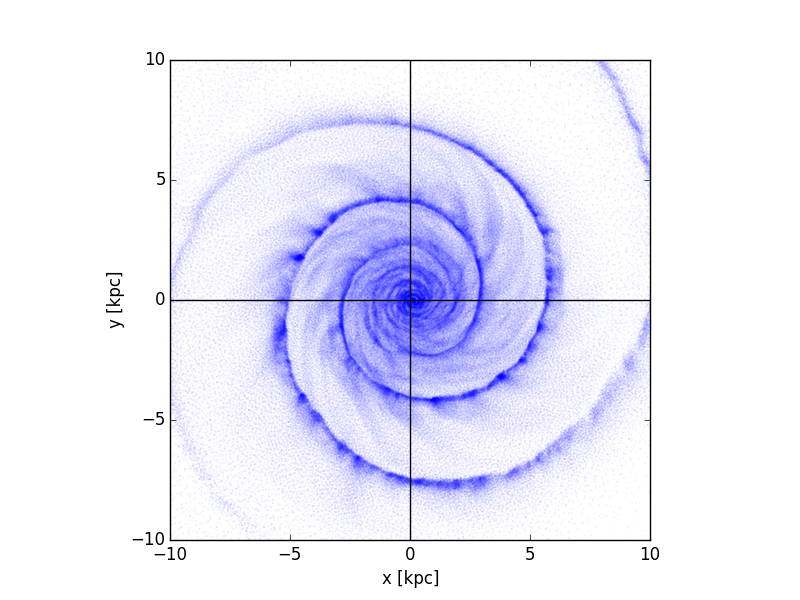}}
\resizebox{.5\hsize}{!}{\includegraphics[trim = 0mm 0mm 0mm 0mm]{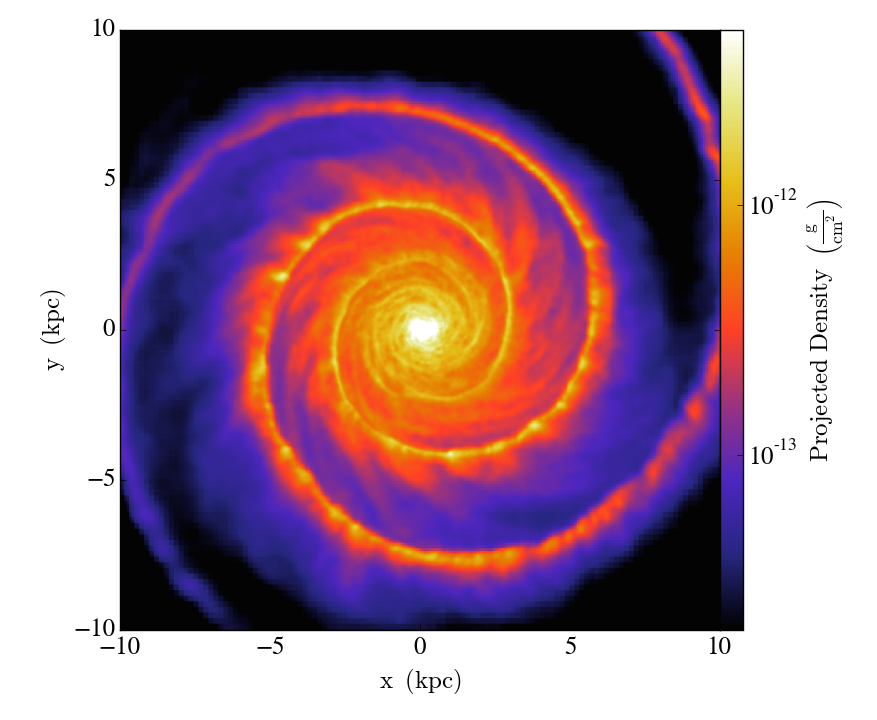}}
\resizebox{.45\hsize}{!}{\includegraphics[trim = 30mm -5mm 10mm 5mm]{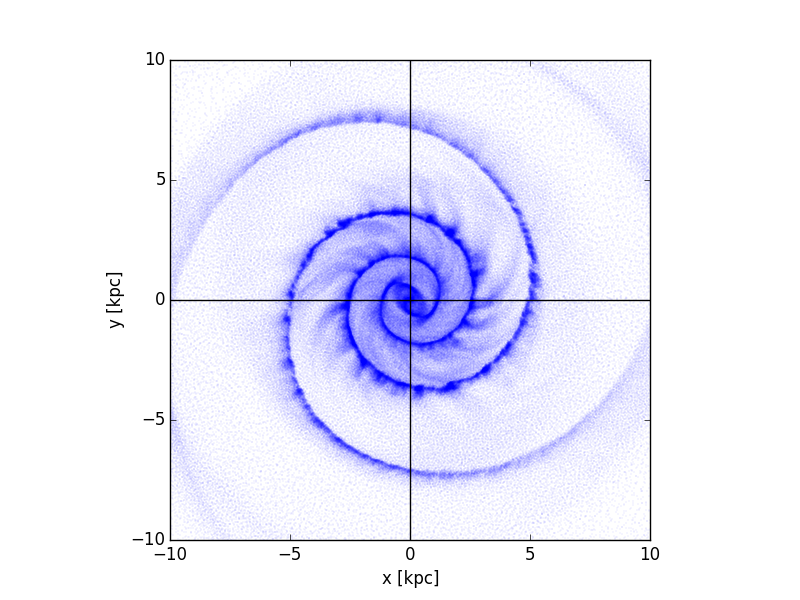}}
\resizebox{.5\hsize}{!}{\includegraphics[trim = 0mm 0mm 0mm 0mm]{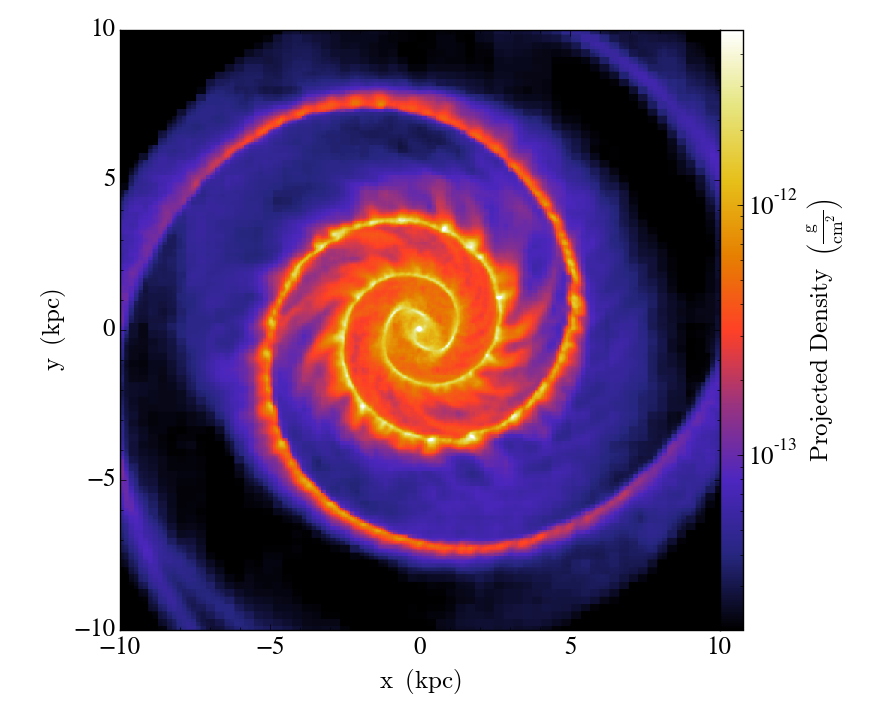}}
\resizebox{.45\hsize}{!}{\includegraphics[trim = 30mm -5mm 10mm 5mm]{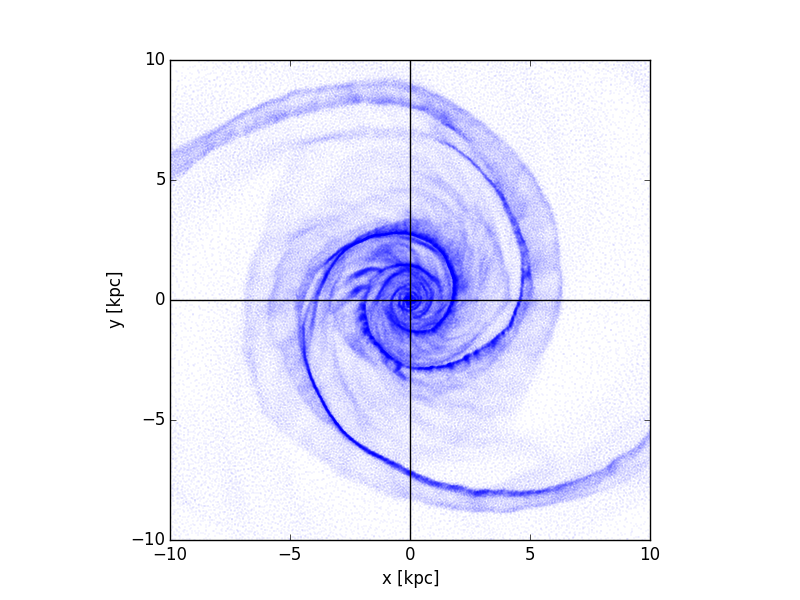}}
\resizebox{.5\hsize}{!}{\includegraphics[trim = 0mm 0mm 0mm 0mm]{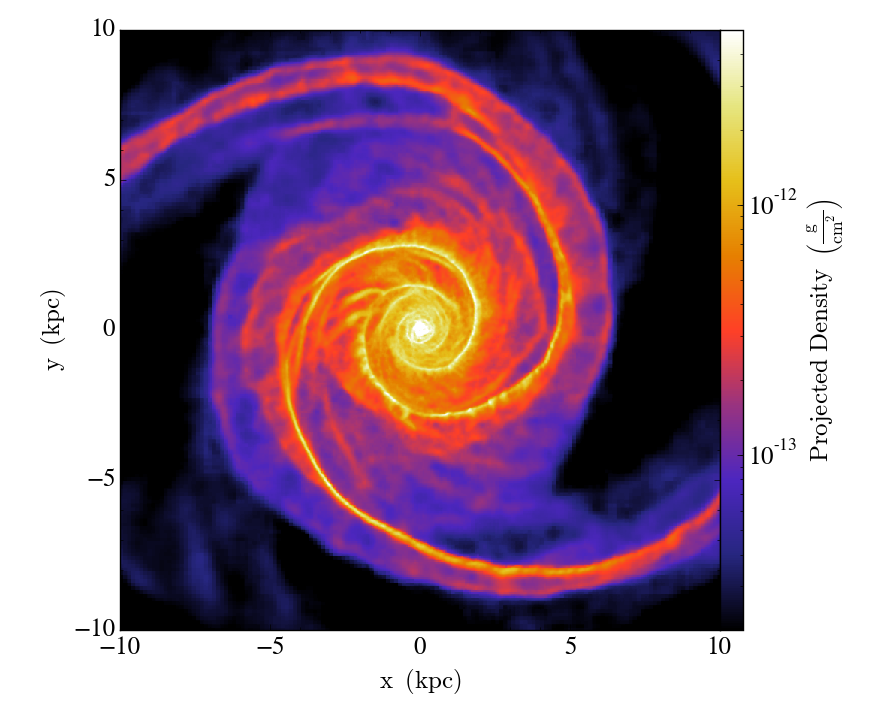}}
  \caption{Gas density render and particle locations in our fiducial tidally driven simulation (top), in a simulation with fixed spiral potential (SDW, middle) and a simulation where a relatively heavy live stellar disc has driven a short-lived two-armed structure (DTR, bottom). Spurs can be seen in the inter-arm regions in all cases, but are much weaker in the live-disc simulation.}
  \label{Spur}
\end{figure}

We find that once the two-armed mode has been established there are noticeable spur-like features present in the galactic gas disc. These are not seen the stellar component. The presence of spurs in external galaxies is most evident in M51, and has also been seen in simulations in the literature  \citep{2003ApJ...596..220C,2006MNRAS.367..873D,2006ApJ...647..997S}. They are relatively easy to produce in simulations with fixed spiral potentials, but somewhat more difficult to reproduce in interactions (seen in \citealt{2010MNRAS.403..625D} and \citealt{2011MNRAS.414.2498S} though not as clear as here), and have remained elusive in simulations of live-stellar discs (e.g. \citealt{2011ApJ...735....1W}). 
In the top row of Fig.\,\ref{Spur} we show a snapshot of our simulation where clear spurring features can be seen (top, including particle positions and density render). In the middle row we show the results of a calculation where the gas is instead exposed to a static stellar potential (SDW model). The gas is exposed to a log-spiral potential of \citet{2002ApJS..142..261C} with a pitch angle chosen to match that of the companion induced spiral (15\arcdeg{}) and a pattern speed of 12\ps{}, which is a medium value of the arms in the perturbed disc. The spiral structures are quite similar, with spurring features existing between the arms in the mid disc region. There are obviously some differences, owing to the variable rotation of the spiral arms and the much more dynamic nature of the stellar component in the perturbed galaxy. The reasons for the existence of these spurs is not fully understood. Possible causes include Kelvin-Helmholtz instabilities \citep{2004MNRAS.349..270W}, orbit crowing as gas passes through a spiral shock \citep{2006MNRAS.367..873D} or vorticity generated at deformed spiral shock fronts \citep{2014ApJ...789...68K} and can be dramatically influenced by magnetic fields and intricacies of the axisymmetric rotation curve \citep{2006ApJ...647..997S}.

For comparison we also show a live-disc calculation without a perturbing companion that has been initialized so that a two-armed structure is produced in isolation (DTR model, bottom row of Fig.\,\ref{Spur}). The arm structure was simply achieved by doubling the stellar disc mass, which encourages the production of low-mode arm formation (see \citealt{2015ApJ...808L...8D,2015MNRAS.449.3911P}). The arms in this calculation are much more transient than the other models, and will soon shear out into new arm structures (due to their material rotation speed). The spirals formed here have very limited spur features, though some do persist in the upper-left quadrant. The lack of spurs in this dynamic spiral is due to their material-like pattern speed and lack of strong spiral shock. The aforementioned possible causes of these spurs all hinge on the passage of gas through the spiral arms regardless of the mechanism. The gas here is coincident with the stellar spiral potential in the mid to outer disc. The central region however has a minor spur feature, which is likely brought on by the shearing out of the spiral arm it resides in rather than passing through a shock.

The existence of these features can be therefore used to discern the origin of spiral structure in observers galaxies, as the two mechanisms shown in Fig.\,\ref{Spur} show spurs, while live disc simulations do not, instead showing more pronounced branches and bifurcations. Future calculations with more realistic ISM physics will help to identify the nature of these spurs, as the warm isothermal calculations presented here are difficult to compare directly to observations.

\subsection{Parameter study}
\label{Res3}

We split our analysis of our parameter sweep into discussion of the variation by mass (Sec. \ref{MassSweep}), then variation by orbital path (Sec. \ref{OrbitSweep}), and finally a brief discussion of the comparative response across all models in terms of strength (Sec. \ref{strength}) migration of material (Sec. \ref{migmat}).

\subsubsection{Varying companion mass}
\label{MassSweep}
One of the purposes of this work is to assess the limiting case of when a companion induces spiral structure, specifically the mass of companion required to form two-armed spiral features. We show the results of six different companion masses; our fiducial calculation, one twice as massive, and four lighter variants. We vary masses by factors of 2 less than our fiducial value ($M_{p,0}=2\times 10^{10}M_\odot$), which equates to approximately 0.7 of the stellar disc mass. The orbit is again in plane to maximize the disc response, increasing the duration of the impulse, and the closest approach to the primary maintained at approximately 20\,kpc.

\begin{figure*}
\centering
\resizebox{0.34\hsize}{!}{\includegraphics[trim = -5mm 0mm 5mm 0mm]{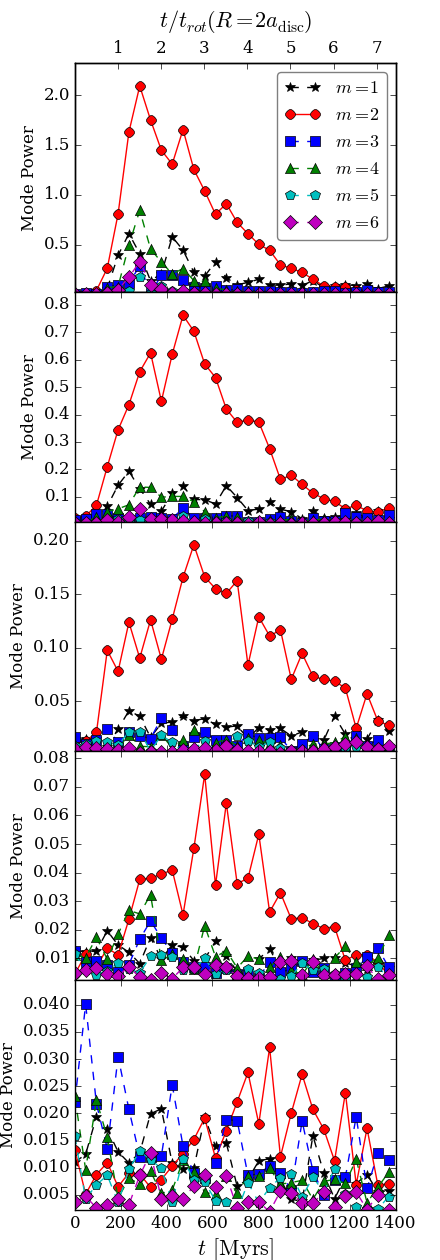}}
\resizebox{0.34\hsize}{!}{\includegraphics[trim = 0mm 0mm 0mm 0mm]{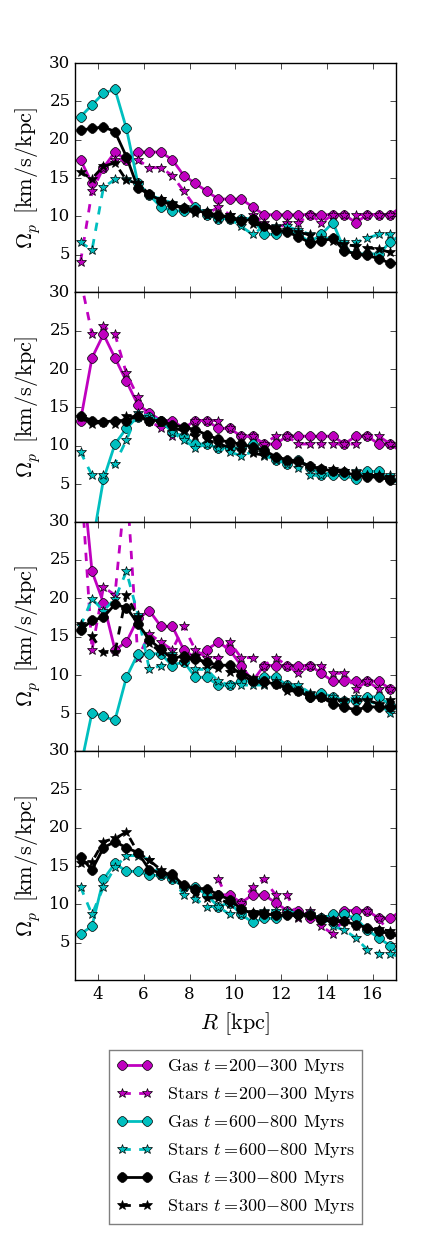}}
\resizebox{0.295\hsize}{!}{\includegraphics[trim = 10mm 4mm 0mm 6mm]{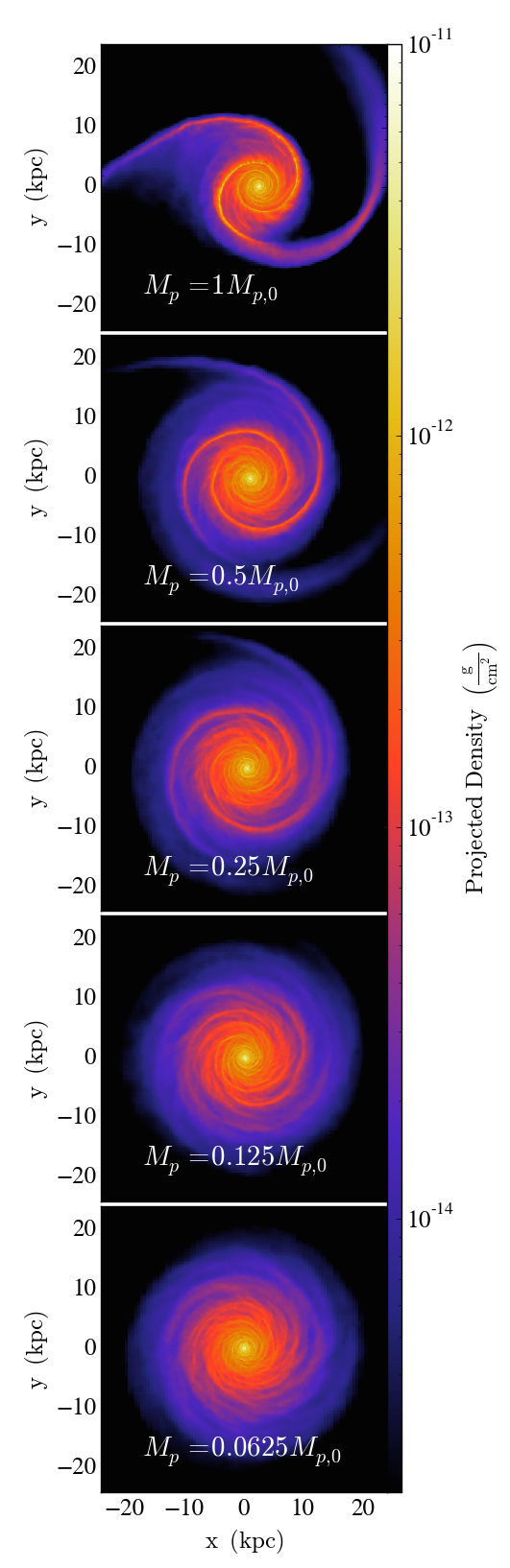}}
  \caption{Fourier arm modes (calculated in $4{\rm kpc}\le R \le 12{\rm kpc}$ range), pattern speeds and gas density render in interactions with various companion masses (decreasing from top). The $m=2$ mode can be clearly seen to drop in power as the companion mass is reduced, reaching barely greater than the ambient noise for the lowest mass case. The pattern speeds show very little variation with mass in either component. Note the spiral arms are too weak to fit a pattern speed in the lowest mass case.}
  \label{MassChange}
\end{figure*}

In Fig.\,\ref{MassChange} we show results for different mass perturbers. In the right hand column we show the gas surface density at the time where the $m=2$ mode is the strongest. The left hand column shows the power of each of the arm modes. These are similar to Fig.\,\ref{ArmModes} but only using the stellar material in a range of $4{\rm kpc} \le R \le 12 {\rm kpc}$. The gas tends to follow a very similar trend except with relatively more low-lying power in the $m=4$ mode and greater noise from other modes in general. The central column shows the pattern speed of the two-armed features at two different epochs after the interaction (early:cyan and late:magenta) in the stars (dashed) and the gas (solid) and over a period of two full rotations (black). Note there is no pattern speed shown for the lowest mass companion because the $m=2$ component was too weak to fit a consistent arm feature to for long enough to calculate a pattern speed, and the early pattern speed for the second to last model was impossible to determine at early times.

The gas renders show a very clear decrease in the disc response to the companion as the mass is decreased. This is also mirrored in the power spectrum, where the $m=2$ mode is seen to be barely any higher than the ambient noise of the remaining modes for the $0.0625M_{p,0}$ mass companion. The overall behaviour of the $m=2$ mode with time also changes with decreasing mass. For the two heavier companions there is a clear sharp increase as the initial bridges and tails are formed, which then slowly decrease to lower levels over the course of a Gyr. With the lighter companions the response is more muted and near-symmetric, in that the $m=2$ mode increases as gradually as it decays, likely due to the lack of a strong bridge adding significant power to the $m=2$ mode.

The pattern speeds show very little change with decreasing companion mass. All masses show a good general agreement with the location of the $\Omega-\kappa/2$ resonance as seen in Fig.\,\ref{PatternSpeed} at later times. The pattern speed is more difficult to fit in the inner disc ($R<5$kpc) where the influence of the random velocities of the bulge is considerable. At the early epoch (magenta lines) the pattern speed is near constant in the outer disc, as documented by \citet{2015ApJ...807...73O}. However, we stress that during this period the interaction is in its early stage and when there is still a clear bridge and tail system in the heavier models. We choose to perform most of our analysis after the bridge is disconnected and when the spiral, avoiding the epochs when the arms are highly asymmetric.

The pitch angles for all the models also follow a similar trend between different mass companions, and we do not show them here. All have an initial steep rise and then gradual decline as in our fiducial model (Fig.\,\ref{Alpha}) though the weaker the model the faster the decrease occurs and the lower the maximum pitch angle initially reached.

\begin{figure}
\centering
\resizebox{1.0\hsize}{!}{\includegraphics[trim = 0mm 0mm 0mm 0mm]{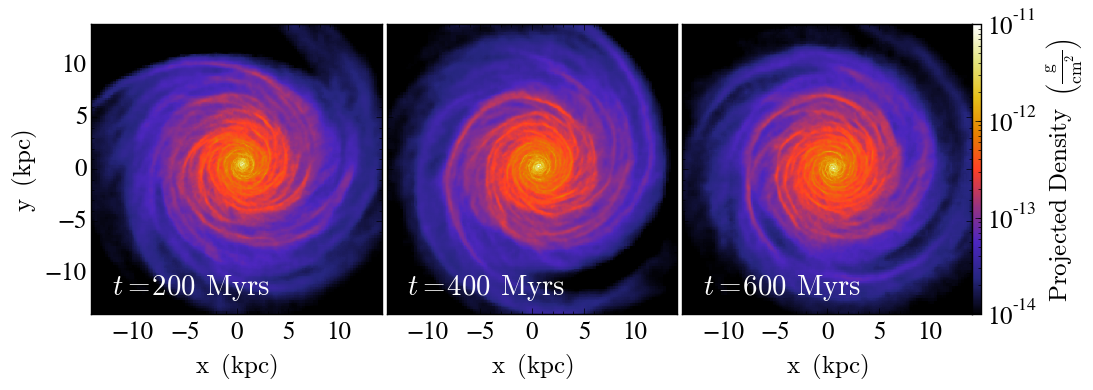}}
\resizebox{1.0\hsize}{!}{\includegraphics[trim = 8mm 0mm -8mm 0mm]{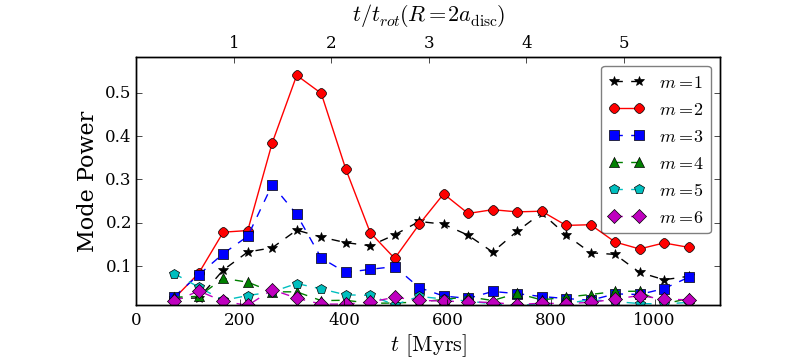}}
  \caption{Response of the model galaxy to the lowest mass companion with a closer periastron passage (model Light4d2). Top panels show the top-down gas response, and the bottom panel shows the evolution of stellar arm response. The Fourier modes are shown in the stars in the range $4{\rm kpc} \le R \le 12 {\rm kpc}$.}
  \label{MassLow}
\end{figure}

As the $0.0625 M_{p,0}$ ($1.25\times 10^{9}M_\odot$) companion induced negligible response in the host galaxy, we ran further calculations to discern whether this is the limit for inducing structure.
We reduced the closest approach distance to the host galaxy, until the disc displayed signs of the interaction (Light4d1). In Fig.\,\ref{MassLow} we show the resulting calculation, where the periastron passage distance is reduced to 12kpc. The top panel shows the density render, and the lower panel the evolution of the Fourier components. By this stage a small SDW is driven in the outer disc, but is fairly weak and accompanied by a growth in the $m=3$ mode initially, and the $m=1$ mode later on. Attempts to increase the amplitude and longevity of the $m=2$ mode by varying the companion properties and orbit resulted in mergers or unperturbed fly-bys. Closer periastron passages resulted in the companion ploughing through the disc, at which point the point-mass approximation breaks down, and the perturber generates a strong $m=1$ mode in the interaction. We therefore find that a companion to stellar disc mass ratio of approximately $f_d \approx 25$ ($\approx 1\times 10^9$ in our calculations) is the limit to significant spiral structure generation, below which it is unlikely spiral features can be induced by a non-merging tidal encounter.
Substructures of this size in galactic haloes are seen in large scale structure simulations (e.g. \citealt{1999ApJ...524L..19M}). While the masses of these subhalos does extend into the $1\times 10^9$ regime, this appears at the tail end of the distributions in some studies \citep{2004MNRAS.348..333D,2004MNRAS.355..819G}. 

\begin{figure}
\centering
\resizebox{1.0\hsize}{!}{\includegraphics[trim = 0mm 0mm 0mm 0mm]{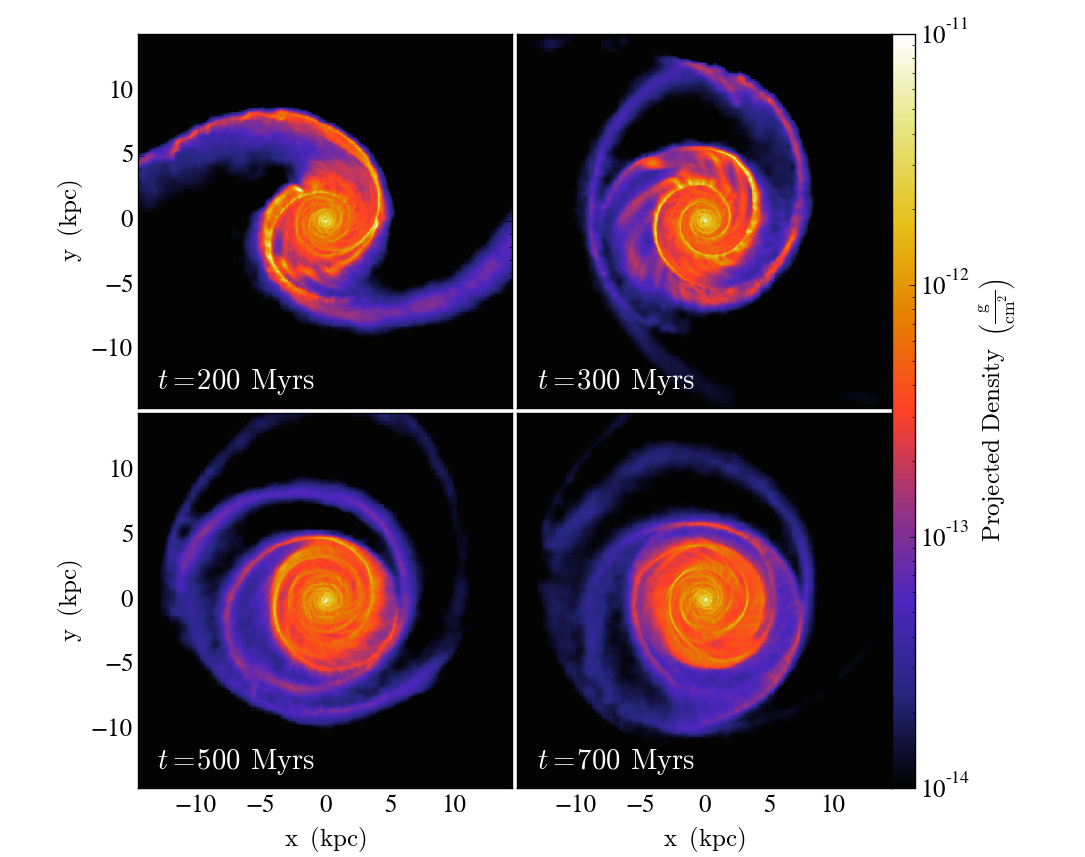}}
  \caption{Time-lapse of the gas density in the interaction between our heavy companion and the host galaxy. The spiral response appears weaker after the interaction compared to the fiducial model (Figure\,\ref{PertLapse}).}
  \label{MassX2}
\end{figure}

For completeness we also include a calculations with a companion with twice the fiducial mass (Heavy1), where the companion mass is now heavier than the stellar disc itself. A time-lapse of the gas response is shown in Fig.\,\ref{MassX2}. The response of the disc in this instance is very strong, which results in a less symmetric spiral structure than the lower mass calculations. An initially very strong bridge-tail feature is formed, which transforms into a one-armed structure shortly after periastron passage (about 200\,Myr). This arm interacts with the more regular inner two-armed features to create some more exotic and irregular ring-like features, and even a leading arm structure in the final panel. Much of the low-density gas has also been radially offset in this interaction compared to the lower mass companions, reducing the effective ``size" of the host galaxy by nearly 2kpc shortly after the interaction compared to the lighter calculations (see Sec.\;\ref{migmat}). Higher mass companions were also tested, but the interaction became more and more destructive and formed short lived spiral arms that quickly formed irregular features.
One benefit of this strongly interacting case is that the stronger tidal forces appear more efficient at driving spiral features in the inner disc. The top-right panel of Fig.\,\ref{MassX2} shows a two-armed feature that persists to $R \approx 2{\rm kpc}$, whereas the fiducial calculation has arms that dissipate by 3-4kpc.

While not explicitly shown here, we also experimented with varying the perturber initial velocity magnitude. The resulting arm structures were very similar to effect seen in varying the companion mass.
The Slow1 calculation (10\kms{} slower initial velocity) drives the same response as doubling the companion mass, and the Fast1 calculation (10\kms{} faster initial velocity) the same as halving the companion mass. This is due to the effect of increasing/decreasing the duration of the impulse experienced by the disc, creating structures that are similar to increasing/decreasing the companion mass. Even the final separation between the host and companion is similar in the Fast1 and Light1 models, reaching approximately 60kpc when the spiral mode is most well defined. The equivalent is true for the Slow1 and Heavy1 calculations, though the orbits are clearly bound in these cases.

\subsubsection{Varying companion orbit}
\label{OrbitSweep}

We perform a limited number of calculations where the perturbing companion is no longer orbiting in the plane of galactic rotation. These include three orbits where the perturber origin is the same, but velocity vector is rotated by 45\arcdeg{}, 90\arcdeg{} (passes over the North Galactic Pole), and 135\arcdeg{} out of plane; Orbit45, Orbit90 and Orbit135. We also perform a single calculation where the companion originates directly above the North Galactic Pole (the Above model), but all other properties of the orbit are same. The orbital path of the Above and Orbit90 models has no azimuthal component at closest approach, and so has no rotation frequency to compare to the galactic disc. The Orbit45 and Orbit135 models have frequencies of $+10$\ps{} and $-10$\ps{} respectively, slightly lower than that of the previous models (13.5\ps{}) and closer to the disc orbital frequency at the same radius. 

\begin{figure}
\centering
\resizebox{1.0\hsize}{!}{\includegraphics[trim = 20mm 0mm 20mm 0mm]{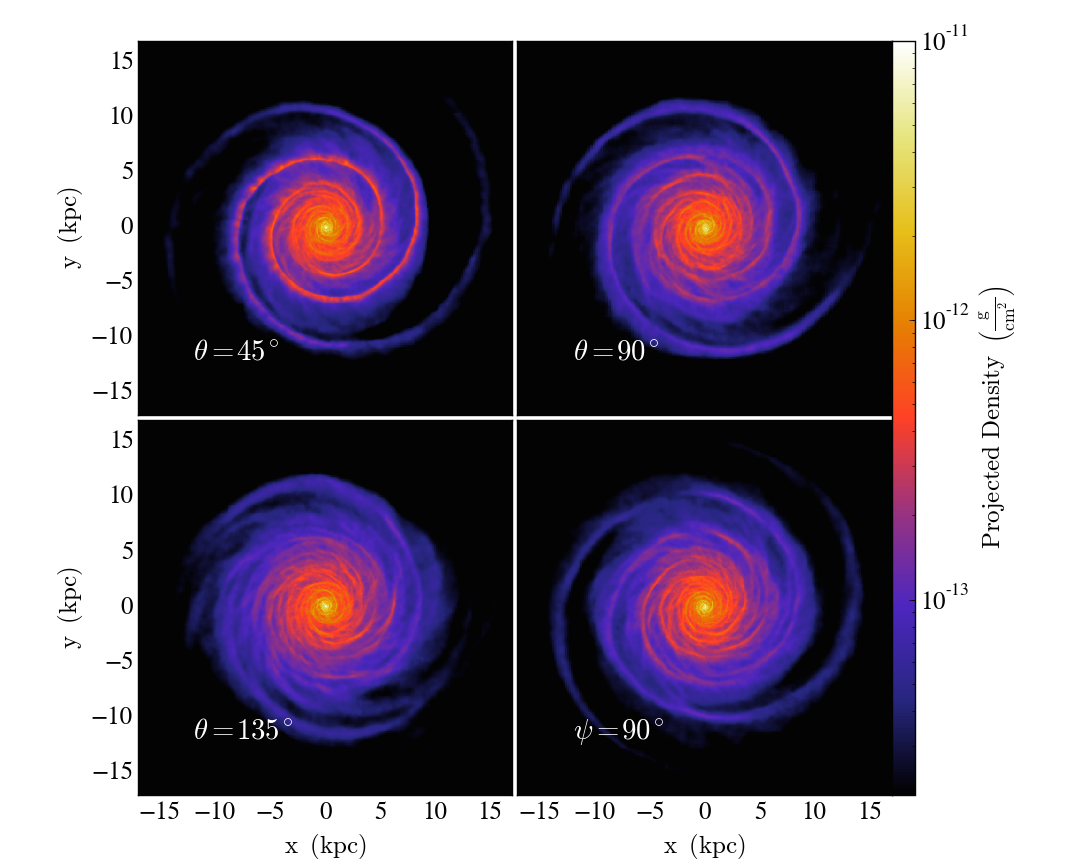}}
  \caption{Gas surface density render for the primary galaxy after interaction with the fiducial mass companion with different orbital paths. Notice that despite the same mass and periastron distance, the resulting arm structure is very different between models.}
  \label{OrbitsRend}
\end{figure}

In Fig.\,\ref{OrbitsRend} we show top-down gas renders of the response of the disc in each of these four calculations  600\,Myr after closest approach. The response of the disc is clearly seen to be reduced the further the companion moves out of plane, with the Orbit45 calculation appearing very similar to $0.5 M_{p,0}$ mass companion from Fig.\,\ref{MassChange}, despite the point of closest approach being unchanged. The retrograde approach (Orbit135) has an extremely diminished effect on the disc, and tests with a completely retrograde in-plane orbit ($\theta=180^\circ$) showed no resulting spiral structure in the disc. Moving the orbital path of the companion of host galaxy's orbital plane therefore gives a similar result as lowering the mass, implying the in-plane and retrograde momentum of the companion is the key quantity.

\begin{figure}
\centering
\resizebox{1.0\hsize}{!}{\includegraphics[trim = 20mm 0mm 10mm 0mm]{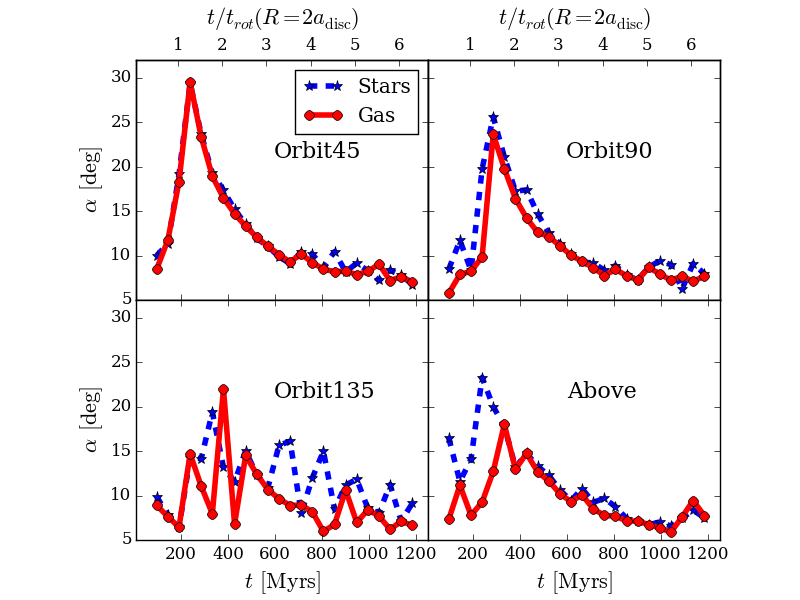}}
  \caption{Evolution of the pitch angle of the $m=2$ arm mode (in stars and gas) induced by companions of our fiducial mass but with different orbital paths.}
  \label{OrbitsAlpha}
\end{figure}

\begin{figure}
\centering
\resizebox{0.9\hsize}{!}{\includegraphics[trim = 15mm 0mm 15mm 10mm]{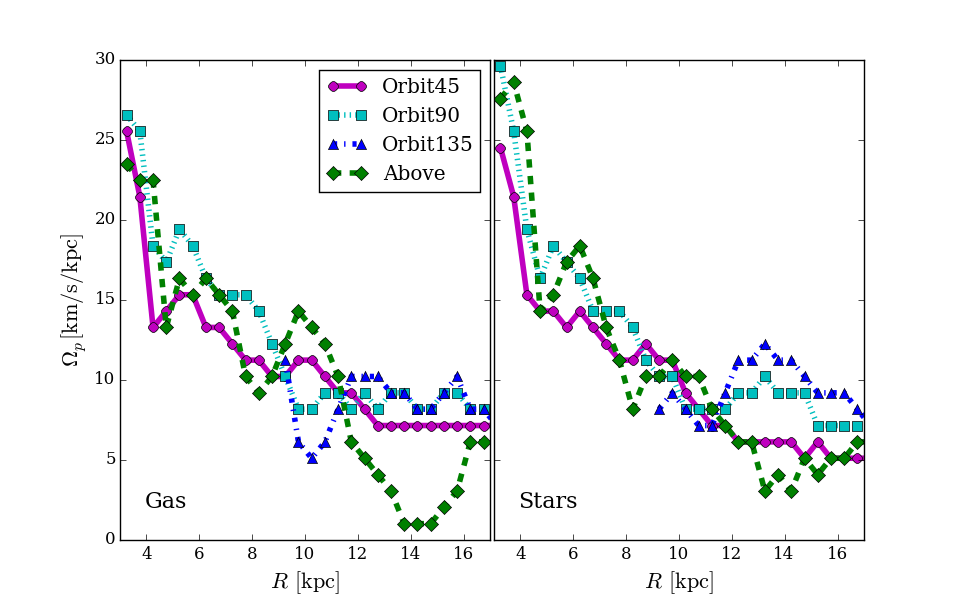}}
  \caption{Pattern speed of $m=2$ features in the galactic disc in the gas (left) and stars (right) driven by companions on different orbital trajectories. Note that the $m=2$ mode is so weak in the Orbit135 model a pattern speed can only be calculated in the outer disc.}
  \label{OrbitsOmega}
\end{figure}

Regarding the properties of the spirals driven in these calculations, we show the evolution of the pitch angle with time in Fig.\,\ref{OrbitsAlpha} and the pattern speed as a function of radius in Fig.\,\ref{OrbitsOmega}. The pattern speeds are calculated at a time where the $m=2$ mode is most prominent. The pitch angles all have very similar behaviour, and start with a decreasing maximum amplitude as the orbits move further out of plane. The rate of decay is similar to that of Fig.\,\ref{Alpha}, dropping to about $4^\circ$ in a Gyr.
The pattern speeds are also similar for all models, where the main difference is seen in the Orbit135 model, whose pattern speed appears flatter than that of the other models. We only show the pattern speed in this model in the $8  {\rm kpc}  \le R \le 17 {\rm kpc}$ range as further within the disc there are negligible arm features to fit to. The near constant pattern speed in this range is similar to that of the $0.125 M_{p,0}$ model in the same range at the earlier epoch, which also has a very weak spiral response (Fig \ref{MassChange}). The stars and the gas again trace very similar pattern speeds for each model.

\begin{figure}
\centering
\resizebox{1.0\hsize}{!}{\includegraphics[trim = 20mm 0mm 20mm 0mm]{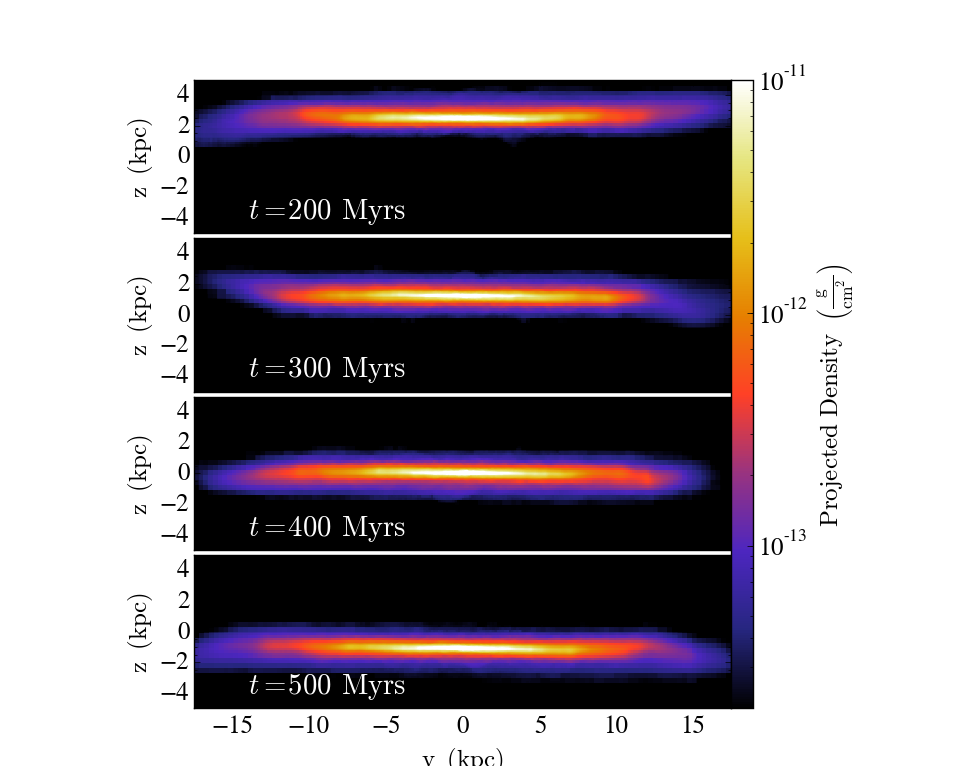}}
  \caption{Vertical projection of gas density at different times in the simulation ``Above" where the companion originates from the North Galactic Pole. A warp can be seen to be induced at the disc edge, which oscillates about the $x-y$ plane before settling back to equilibrium.}
  \label{Warps}
\end{figure}

In Fig.\,\ref{Warps} we show edge-on views of the gas disc in the calculation where the companion originates above the disc (bottom right Fig.\,\ref{OrbitsRend}). Panels show different times after periastron passage ($t=0$). A clear warp feature can be seen in the outer edge of the disc ($12{\rm kpc}\le R \le 20{\rm kpc}$), which oscillates about the $x-y$ plane after the passage of the companion and has stabilized after approximately 500\,Myr. Galactic warps are not uncommon in external galaxies (e.g. ESO 510-G13) and our own Milky Way \citep{2006ApJ...643..881L,2006ApJ...641L..33W}. The warp of the Milky Way is seen to extend to about 1kpc at $R\approx 15{\rm kpc}$, which is very similar to what is seen in Fig.\,\ref{Warps} \citep{doi:10.1146/annurev-astro-082708-101823}. Interestingly for a warp of this scale there is very little spiral structure driven in the disc, especially inside $R\le10${\rm kpc}. This suggests that whatever process induces warps in galactic discs need not necessarily be responsible for observed spiral structure. For example, in the context of the Milky Way, if the Magellanic companions are responsible for the Galactic warp then the spiral structure itself may be driven by a different mechanism. 

\subsubsection{Quantifying the strength of different interaction scenarios}
\label{strength}

\begin{figure*}
\centering
\resizebox{0.48\hsize}{!}{\includegraphics[trim = 0mm 0mm 0mm 0mm]{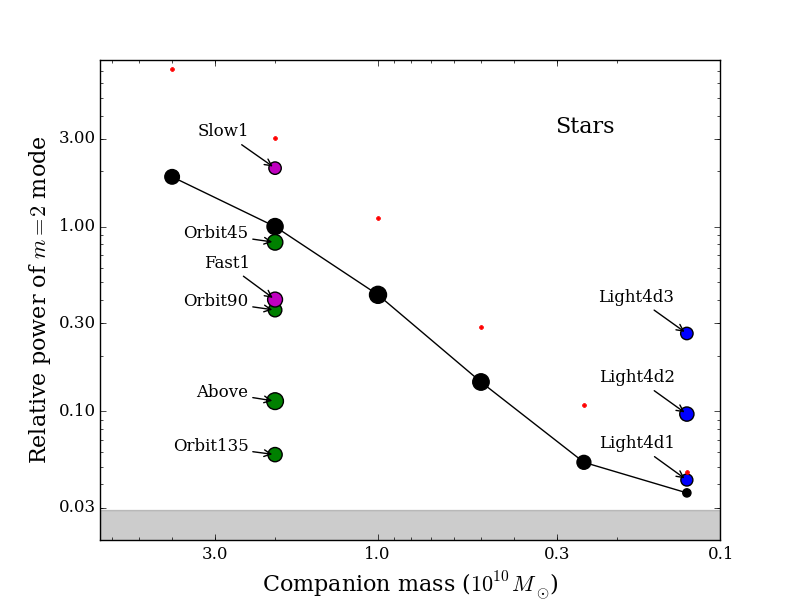}}
\resizebox{0.48\hsize}{!}{\includegraphics[trim = 0mm 0mm 0mm 0mm]{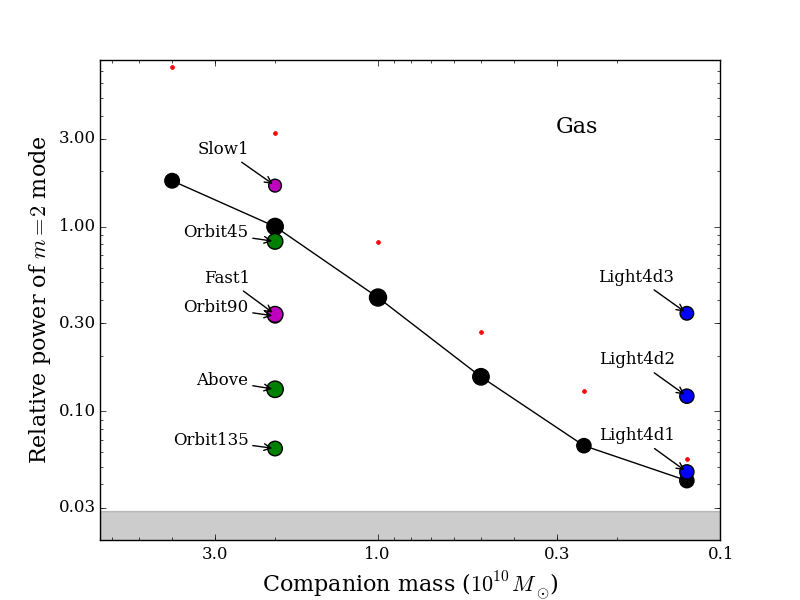}}
  \caption{Comparison of the $m=2$ response in the calculations presented here. The main mass sweep is shown by the black points, the different orbital paths in magenta, different orbital paths in green and low mass companions with different closest approaches in blue. The large circles show the average power of the $m=2$ mode relative to that of the fiducial model. The red points show the maximum for the varying mass models, and the grey region shows the power of the $m=2$ mode before the interaction (the approximate noise level).}
  \label{PowComp}
\end{figure*}
 
Fig.\,\ref{PowComp} shows the arm response for the majority of the calculations presented here. The $y$-axis shows the power of the $m=2$ mode relative to the power of arms in our fiducial calculation, averaged over 1Gyr after the interaction. The black points show different mass companions, with the red points showing the maximum power for these runs, rather than the average. The blue points show the runs with the lightest mass companion for varying closest approach distances. The green points show different orbital paths, and the magenta points show calculations with the $\pm10$\kms{} velocity boost. The size of the points is proportional to the duration for which the $m=2$ mode dominates the full spectrum. The stellar response is shown in the left, and gas in the right, with masses and mode powers shown in log-space. It is immediately clear that there is a near-linear drop in response with decreasing companion mass at fixed closest approach (20kpc), with the lightest companion inducing a response barely stronger than the noise level and being relatively short-lived (as seen in the bottom-left of Fig.\,\ref{MassChange}). Decreasing the closest approach distance then increases the response, though the 16kpc approach (Light4d1) shows little additional power. The Light4d2 (12kpc distance) model shows a stronger response similar to a companion of $\times4$ the mass, and the Light4d3 model stronger still. This latter model however is highly destructive to the host galaxy, carving a great swath through the gas disc and resulting in a merger scenario. 

The out-of-plane companions can be seen to drive a decreasing response in the disc, though being 45\arcdeg{} out of plane seems to make only a minor difference. Moving 90\arcdeg{} out-of-plane is similar to reducing an in-plane companion mass by 1/2, and 135\arcdeg{} similar to reducing by 1/8.

The strength of the interaction can be characterized by the dimensionless parameter \citep{1991A&A...244...52E}:
\begin{equation}
S =  \left(\frac{R_{\rm enc}}{d} \right)^3  \frac{\Delta T}{T}\frac{M_p}{M_{\rm tot}(R<R_{\rm enc})}
\label{Seq}
\end{equation}
where $M_p$ is the companion mass and $d$ is the distance of closest approach. $R_{\rm enc}$ is a characteristic distance of the galaxy (taken here to be 20kpc, the truncation distance of the stellar and gaseous disc) and $M_{\rm tot}(R<R_{\rm enc})$ is the mass of all the host galaxy components within this radius. $\Delta T$ is the time for the perturber to move 1 radian at closest approach, and $T$ is the time for stars at $R_{\rm enc}$ to move 1 radian in orbit around the galactic centre.
This $S$ parameter provides information on the tidal strength of the interaction and is the force experienced by material in the outer edge of the disc over a duration $\Delta T$ as a fraction of the circular momentum in the galactic orbit at this point. This offers a method of characterizing the strength of the interaction while taking into account the velocity information. 

\begin{figure}
\centering
\resizebox{1.0\hsize}{!}{\includegraphics[trim = 0mm 10mm 0mm 0mm]{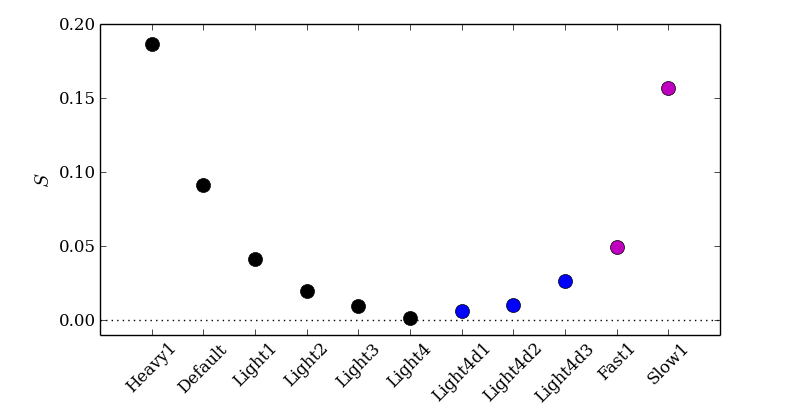}}
  \caption{The dimensionless strength parameter, $S$ (Eq.\,\ref{Seq}), for each of our in-plane interactions with varying orbital properties. Colours are the same as those in Fig.\,\ref{PowComp}. The dotted line traces the $S=0$ limit.}
  \label{Spow}
\end{figure}

We show the values of $S$ for our in-plane interactions in Fig.\,\ref{Spow}, using the same colours for points as Fig.\,\ref{PowComp} for reference. As with the $m=2$ mode analysis there is a clear trend with our models to have a decreasing tidal force with decreasing companion mass. The value of $S$ for our fiducial calculation is similar to those used in the literature for interacting galaxies, however our minimum spiral case is substantially lower than that seen in the literature ($S\approx0.01$ for the Light4d2 calculation). For example, \citet{2015ApJ...807...73O} find a tidal strength of $S\lesssim 0.065$ is required to form at least a tidal tail, however their models do not explore below this value to find the no-spiral case. \citet{1991A&A...244...52E} look at much weaker interactions, and find spirals can be induced in interactions with strengths of $S\approx0.02$, though they are mostly concerned with ocular/bar shaped structures.

\subsubsection{Migration of material}
\label{migmat}

\begin{figure*}
\centering
\resizebox{1.0\hsize}{!}{\includegraphics[trim = 20mm 20mm 10mm 0mm]{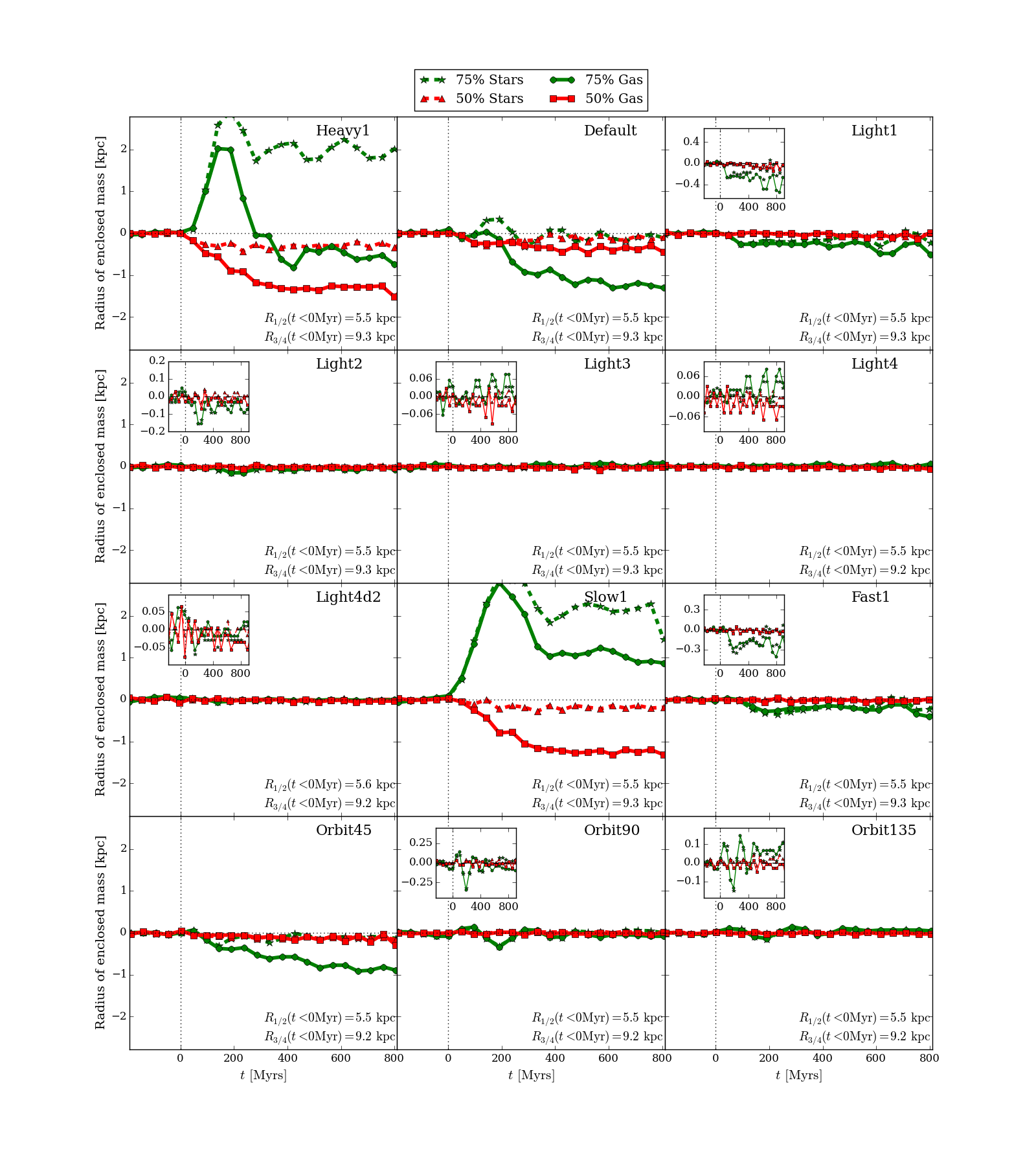}}
  \caption{The evolution of the radius which encloses half ($R_{1/2}$, red lines) or three-quarters ($R_{3/4}$, green lines) of the total mass of stars (dashed lines) and gas in each of our main models (solid lines). The 0\,Myr time corresponds to the time of closest approach of the companion. Radii are measured as offsets from the average value prior to the companion passage, with values given in the lower right corner. All radii are shown to the same scale, with the inserts showing a zoom in of the same data for simulations where the response is small.}
  \label{SDall}
\end{figure*}

In Fig.\,\ref{SDall} we show the radial migration of gas and stars in our calculations. The $y$-axis shows the radius that encompasses either half ($R_{1/2}$, red) or three quarters ($R_{3/4}$, green) of the total gas or stellar mass of the galaxy as a function of time. The values for the radius are shown as offsets to the average value before the interaction, shown in the bottom right corner of each panel. All models are shown to the same scale, and for values whose response is very weak the same data is shown in the zoomed in insert. The Above model is not shown as it shows very similar features to the Orbit90 and Orbit135 calculations. Gas is shown by the circles and solid lines, and stars by the stars and dashed lines. In most cases the material is seen to be migrating inwards, shown in the radii entering the negative region. The infall of gas does not continue for a long period, and in the Default and Heavy1 models levels out after 200\,Myr. The spiral arms are continuously winding after this period however (Figs\,\ref{PertLapse} and \ref{Alpha}) so the motion of material inwards is not a result of changes in the arm structure, but rather the strong tidal force of the companion's passage and lasting 200-400\,Myr.

\begin{figure*}
\centering
\resizebox{1.0\hsize}{!}{\includegraphics[trim = 20mm 20mm 10mm 10mm]{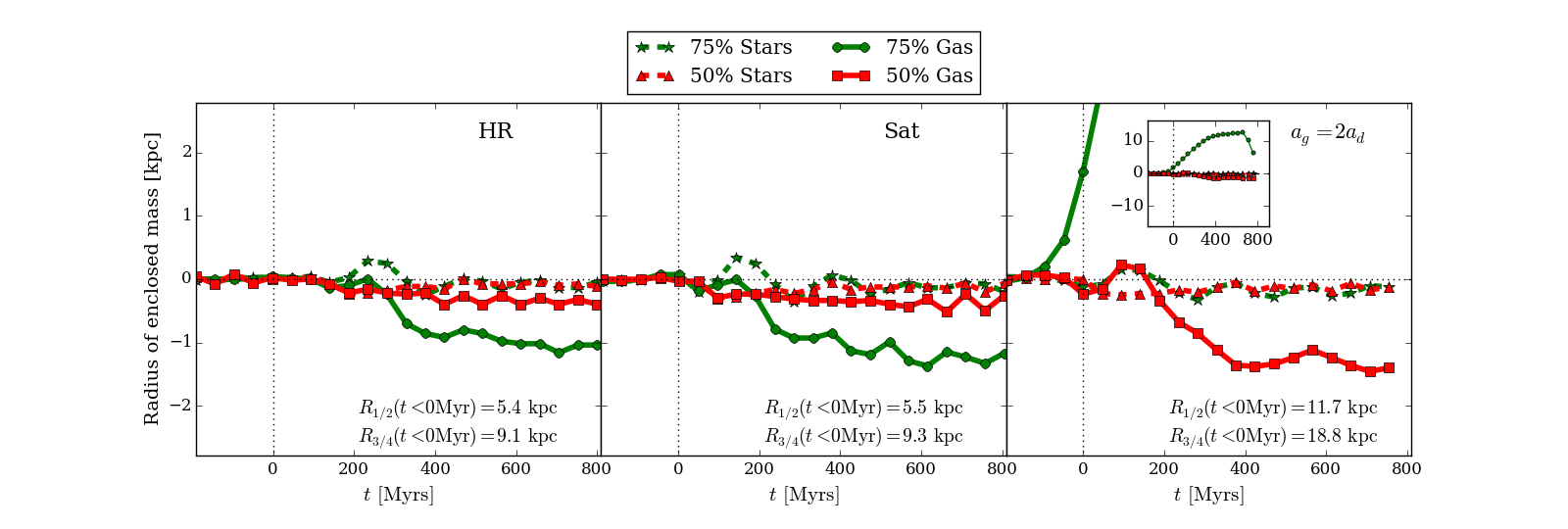}}
  \caption{As Fig. \ref{SDall} but showing the variations on the fiducial calculation. Left: the higher resolution run, centre: the calculation with a resolved companion and right: the calculation with the extended gas disc. The insert in the latter shows a zoom out of the main plot that encompasses the full extent of $R_{3/4}$.}
  \label{SDex}
\end{figure*}

In the strongest interactions (Heavy1 and Slow1) there is also a motion of gas away from the galaxy, shown by the increase of $R_{3/4}$ with a simultaneous infall in the inner disc. In this instance the gas is being stripped from the host to a significant degree. The weakest interactions show no significant radial migration of gas, with $R_{1/2}$  and $R_{3/4}$ moving by only 50pc. In these instances the power of the different arm modes in Fig.\,\ref{MassChange} is a better indicator of the disc response. This is highlighted by the Light3 and Light4 models, as while they have a near-identical behaviour in $R_{1/2}$ and $R_{3/4}$, there is a much clearer $m=2$ component in Fig.\,\ref{MassChange} for Light3.

The gas migration for the Light1, Fast1 and Orbit45 models show a similar trend, highlighting that the resulting tidal forces are similar in each case. Similarly the Heavy1 and Slow1 models show a similar trend, through the slower companion has a stronger effect on the migration, with $R_{3/4}$  showing an increase for a longer time than the Heavy1 calculation.

The gas and stellar material displays a very similar behaviour for most models, especially in the lower strength interactions, where $R_{1/2}$ and $R_{3/4}$ are near indistinguishable. In some cases the gas appears stronger effected by companion (Default, Orbit45, Slow1, Light1) and experiences a greater net motion inwards. The main difference is seen in the Heavy1 interaction, where the stars have an increase in $R_{3/4}$, whereas the gas experiences a rise then a drop back inwards. In the Slow1 interaction, which appears slightly stronger than the Heavy1, the gas and stars both maintain an increase in $R_{3/4}$. This implies the stars are easier to be dragged out of the disc compared to the gas, whereas the gas requires a stronger interaction to be pulled out but conversely will more readily fall inwards than the stars (seen $R_{1/2}$ in Slow1 and Heavy1).

The lack of a strong stripping of gas in these calculations may seem at odds with the paradigm that interaction events should strip spirals of their gas \citep{1999MNRAS.308..947A,2004IAUS..217..440C,2004cgpc.symp..305V}. However, these mechanisms are usually strong interactions or include some dense inter-cluster medium to efficiently strip the outer gas disc. In the models in Fig.\,\ref{SDall} the encounter is grazing, and is only marginally effective at capturing gas in the stronger interactions. Indeed in the tidal interactions that are efficient at gas stripping are usually strong enough that the system results in a merger \citep{2004IAUS..217..440C} which is not the case here. The right hand panel of Fig.\,\ref{SDex} shows the migration of material in a calculation with an extended gas disc. This calculation shows a very different evolution for the gas disc, with the outer material being clearly stripped away from the galaxy (the sharp increase in $R_{3/4}$) and the inner disc falling into the galactic centre. This is due to the material being ram-pressure stripped by the companion which now ploughs through the extended gas disc, whereas in the original calculations the companion grazed the disc-edge.

Fig. \ref{SDex} also shows the evolution for $R_{1/2}$ and $R_{3/4}$ for the simulations with a higher resolution and a resolved companion. Behaviour is very similar to the fiducial calculation shown in Fig. \ref{SDall} in both cases.

\section{Conclusions}
\label{conclusions}

We have performed a set of $N$-body and hydrodynamical simulations of galaxy interactions to better understand the relation between resulting stellar and gas structures, and determine the limiting mass case for spiral generation. We find that the spiral structures are very similar in the gas and stellar components. The arm numbers, pitch angles and pattern speeds are all very similar, with subtle differences such as a more noisy gas power spectrum. We find a small but noticeable offset between the gas and stellar spiral arms, whereas in spiral potentials there is a significant offset and no offset seen in isolated $N$-body spiral arms \citep{2015PASJ...67L...4B}. The pattern speeds of the spiral arms trace the $\Omega-\kappa/2$ curves of material, thus behaving as slowly winding density waves. As gas passes through the spiral potential it exhibits spurring features, which are also not seen in isolated $N$-body simulations. 
The existence of spurs, gas-spiral offsets, and the radial dependence of the pattern speed therefore presents possible tests of the nature spiral arms in nature, which could be paramount for explaining the existence of grand-design two-armed spirals.

We find it possible to quantify the strength of the interactions by either the relative power of the Fourier modes, or a dimensionless strengths parameter \citep{1991A&A...244...52E} that also includes velocity information. Moving the interaction out-of-plane is similar to significantly reducing the mass of the companion, and angling out of plane by 90\arcdeg{} and 135\arcdeg{} produces a similar response to reducing the companion mass by 1/2 and 1/8  respectively. We find a strength parameter of the order of $S\approx0.01$ can induce some spiral features, noticeably lower than found in previous studies.

For our calculations we find a minimum mass limit of approximately $1\times 10^{9}M_\odot$ (equivalent to 4\% of the stellar disc mass), with a closest approach of 12kpc and  whereby spiral structure can barely be generated without a merger scenario. This is within the range of dark matter subhaloes, (e.g. \citealt{2004MNRAS.348..333D,2015arXiv150605537Z}) suggesting small dark matter structures can drive at least a portion of spirals seen in nature.

The next step is to use the work presented here as a foundation to investigate the ability of interactions to reproduce unbarred two-armed spirals in nature. To better match real external galactic gas structure it will be prudent to include the effects of star formation, feedback and ISM heating/cooling.
As the Milky Way presents an ideal nearby test-case of a spiral galaxy, it is highly desirable to understand its spiral features. 
The minimum mass we find above is well below the mass of the LMC ($6-20\times  10^{9}M_{\odot}$; \citealt{1997ApJ...488L.129K}). 
We aim to expand upon our previous studies of the morphology of the Milky Way by assessing whether tidally induced spiral structure can induce the correct observational analogous seen from Earth, as opposed to steady SDW \citep{2014arXiv1406.4150P} or DTR spiral structures \citep{2015MNRAS.449.3911P}.

\section*{Acknowledgements}
We would like to thank the referee for their detailed and informative report which greatly improved this paper. 
ARP is currently supported by the MEXT grant for the Tenure Track System of EJT.
Numerical computations were [in part] carried out on Cray XC30 at Center for Computational Astrophysics, National Astronomical Observatory of Japan and the GPC supercomputer at the SciNet HPC Consortium \citep{2010JPhCS.256a2026L}. SciNet is funded by: the Canada Foundation for Innovation under the auspices of Compute Canada; the Government of Ontario; Ontario Research Fund - Research Excellence; and the University of Toronto.
Figures showing SPH density were rendered using the freely available \textsc{yt} toolkit \citep{2011ApJS..192....9T}. The authors would like to thank researchers at NAOJ, McMaster University, IMPU and ELSI for useful discussions.

\bibliographystyle{mnras}
\bibliography{Pettitt_tidalarms.bbl}

\appendix
\section[]{Additional arm models and resolution}
\label{Appx1}

\begin{figure}
\centering
\resizebox{1.\hsize}{!}{\includegraphics[trim = 10mm 10mm 20mm 10mm]{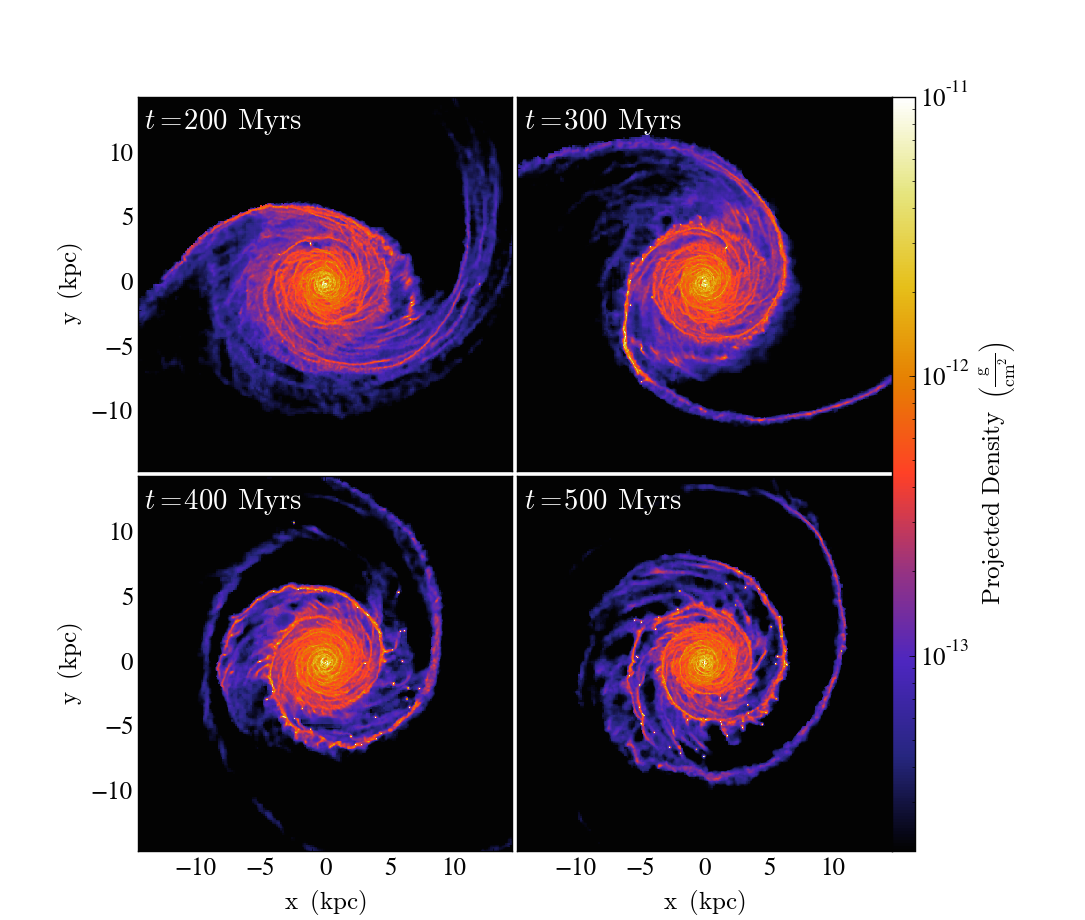}}
  \caption{Top-down surface density of gas in the DefaultCld simulation at four different times after the passage of the companion. Gas here is isothermal at 1000K as opposed to 10000K for the other calculations.}
  \label{ColdP}
\end{figure}

A time-lapse of the DefaultCld calculation is shown in Fig.\,\ref{ColdP} where gas is at a temperature of 1000K instead of 10000K and half the normal surface density. The behaviour of the gas is morphologically similar to the warmer calculations, but displaying a greater degree of structure in the inter-arm regions. Spurring effects are still evident, but the arm features are more fragmented due to the collapse of gas into cloud-like features. These bright, very dense clumps would be the sites of star formation if the relevant physics were active, but in this simulation collapse is simply halted on the gravitational softening scale. High densities are reached initially in the tidal-arm (rather than the bridge-arm) as seen in the 300Myr timestamp which then creates the first population of dense clumps. These dense clumps then pass into the interam region (seen in $x>0$ at 400Myr) at which point the bridge arm has also produced its own high-density clumps. It would be interesting to see if this asymmetric collapse were evident in calculations with star formation and feedback, as then could be used to identify tidal interactions in nature. We leave this to a future study.

\begin{figure}
\centering
\resizebox{0.6\hsize}{!}{\includegraphics[trim = 0mm 0mm 0mm 0mm]{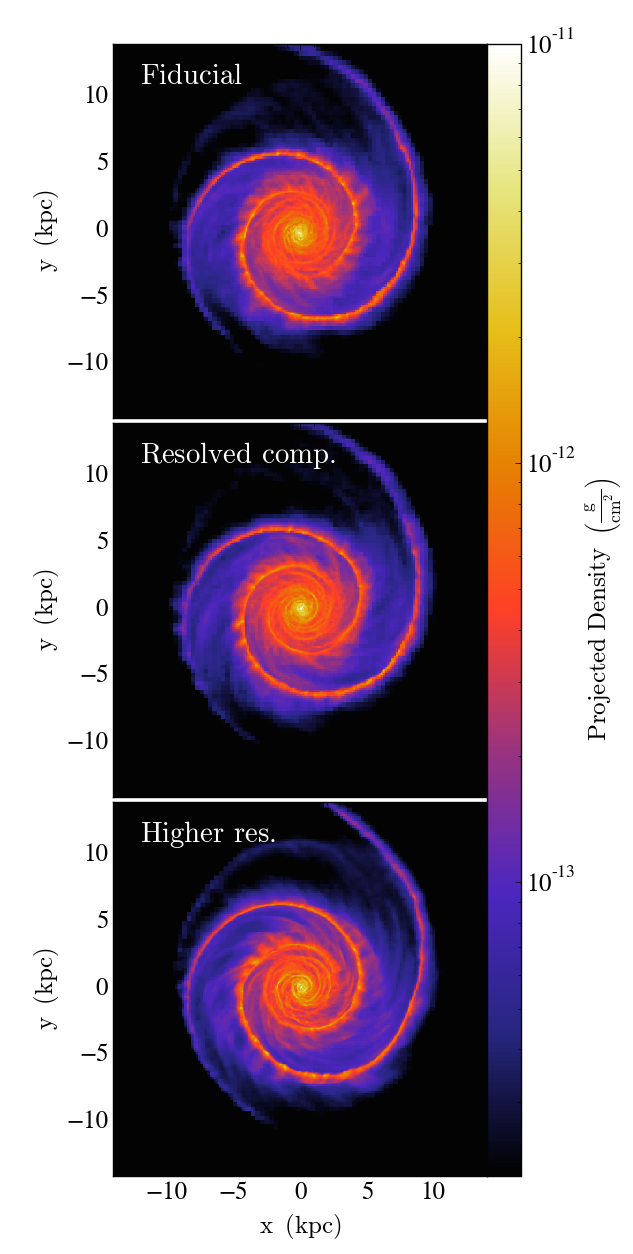}}
  \caption{Density render of gas in simulations with differing mass resolutions. In the top plot is our fiducial set up, where the companion is a point mass. In the centre the companion is instead resolved into 10000 individual particles. In the lower plot the resolution of the main galaxy has been increased x3 the standard value (giving a total particle number of 3 million). The global morphology is very similar in each case.}
  \label{Models}
\end{figure}

\begin{figure*}
\centering
\resizebox{.22\hsize}{!}{\includegraphics[trim = 2mm 0mm -2mm 0mm]{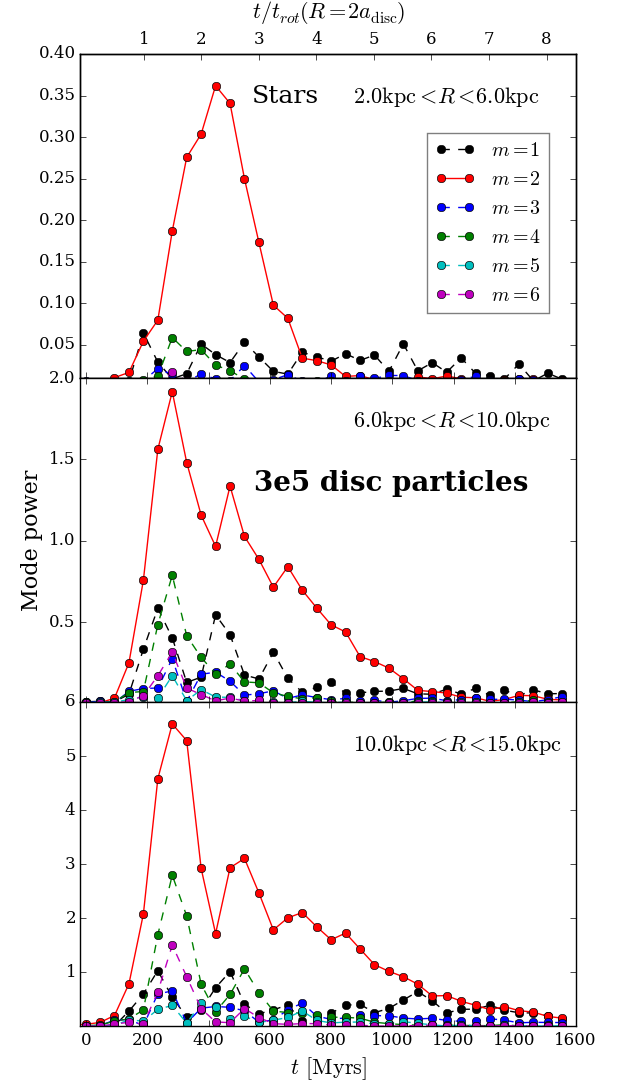}}
\resizebox{.22\hsize}{!}{\includegraphics[trim = 2mm 0mm -2mm 0mm]{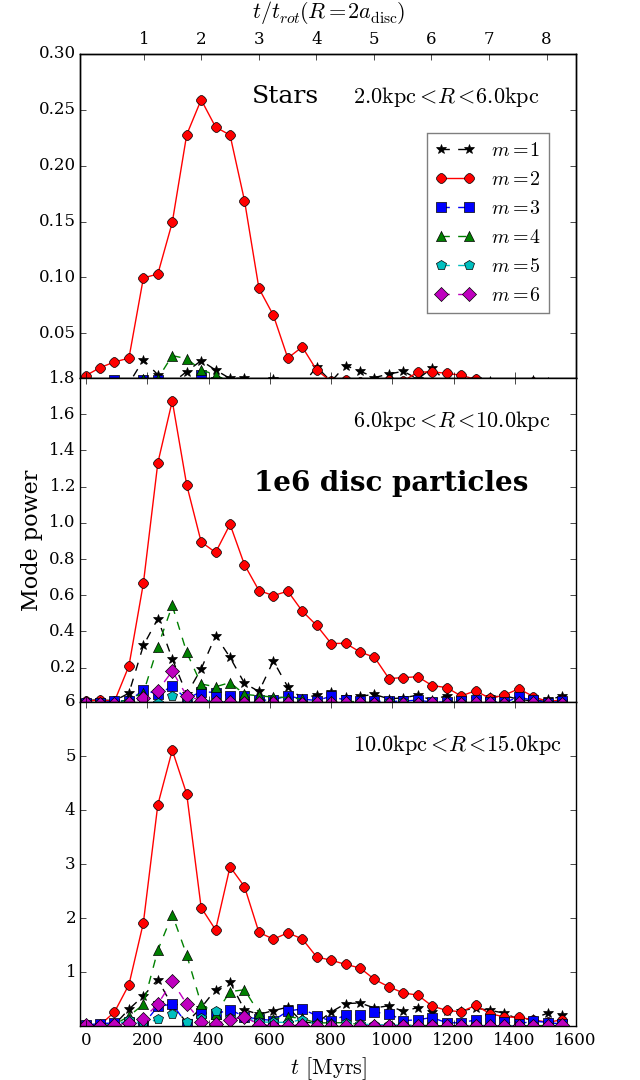}}
\resizebox{.22\hsize}{!}{\includegraphics[trim = 2mm 0mm -2mm 0mm]{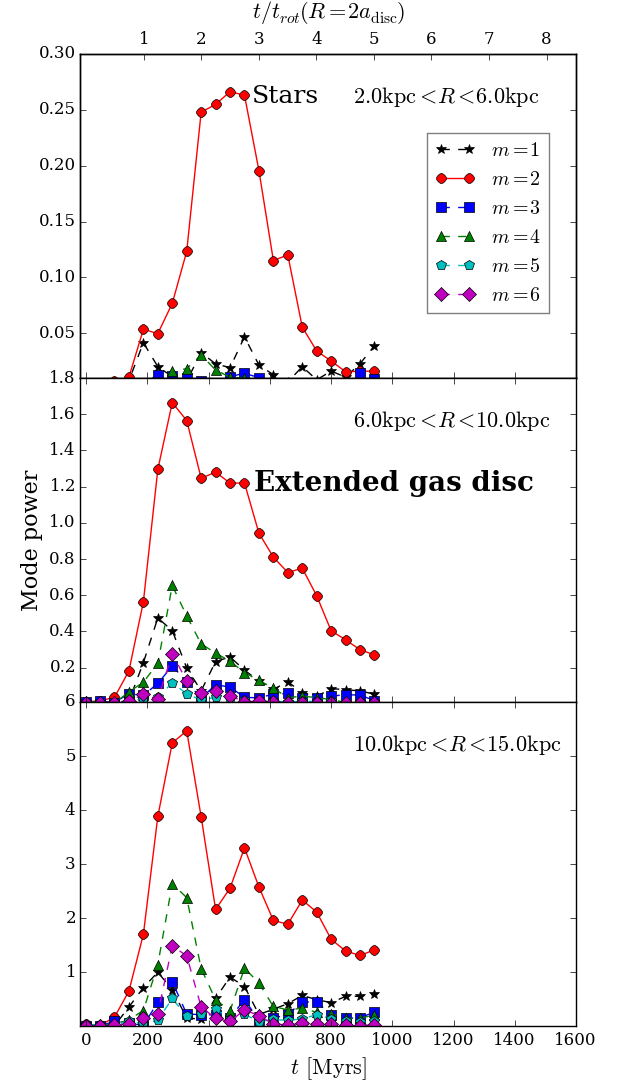}}
\resizebox{.22\hsize}{!}{\includegraphics[trim = 2mm 0mm -2mm 0mm]{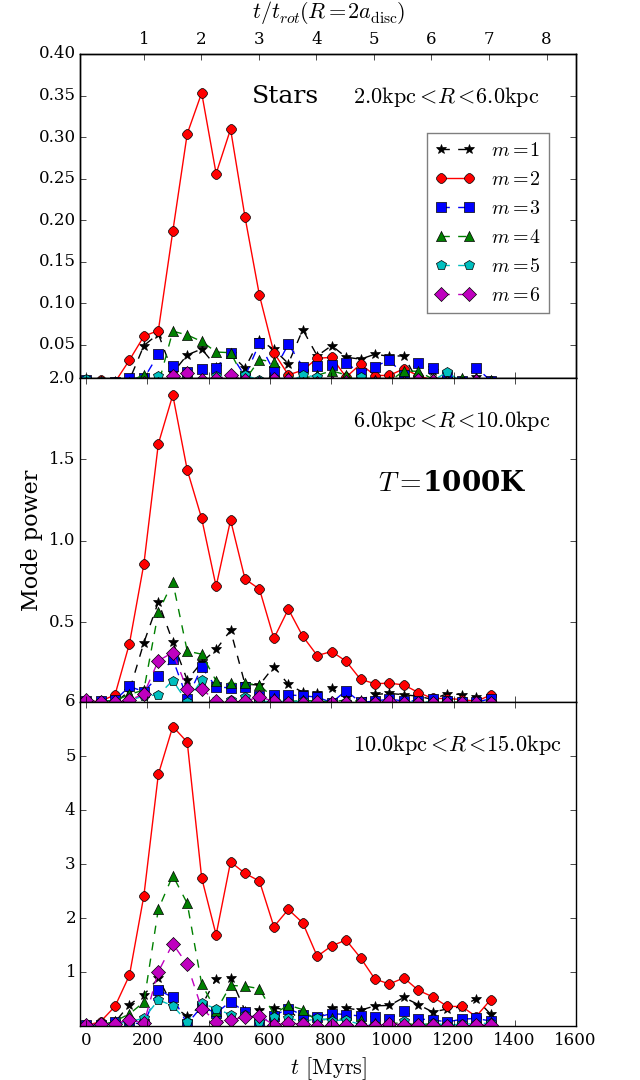}}
\resizebox{.22\hsize}{!}{\includegraphics[trim = 0mm 0mm 0mm 0mm]{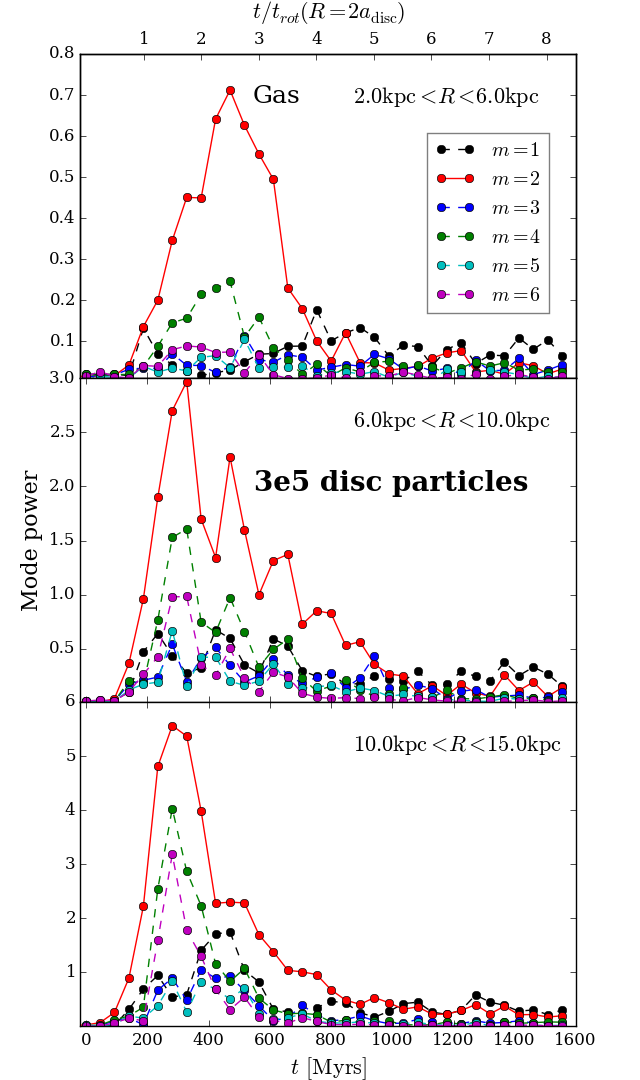}}
\resizebox{.22\hsize}{!}{\includegraphics[trim = 0mm 0mm 0mm 0mm]{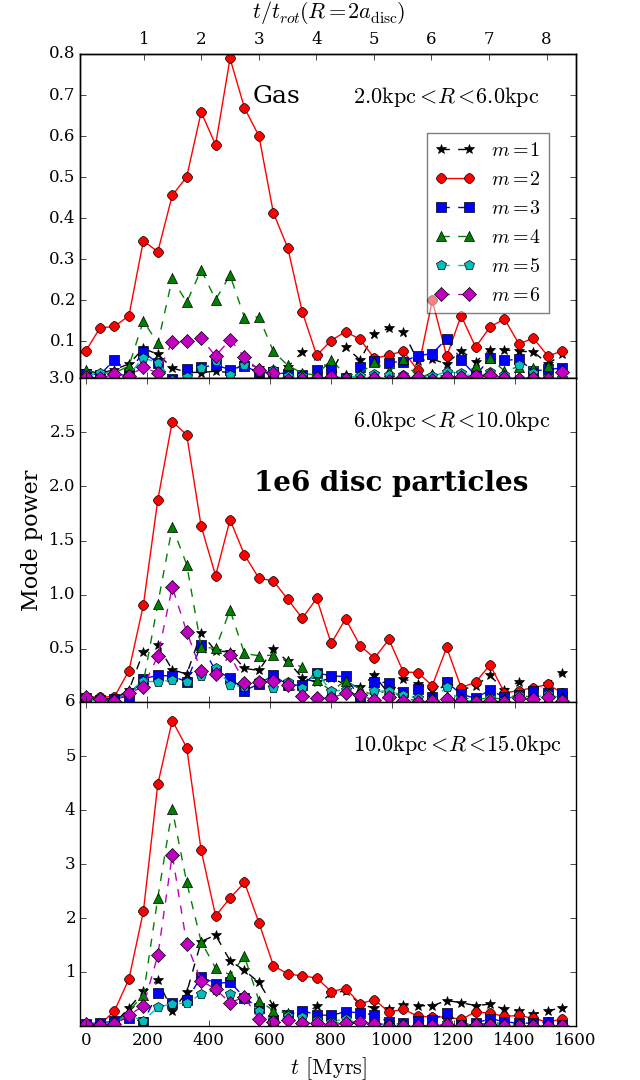}}
\resizebox{.22\hsize}{!}{\includegraphics[trim = 0mm 0mm 0mm 0mm]{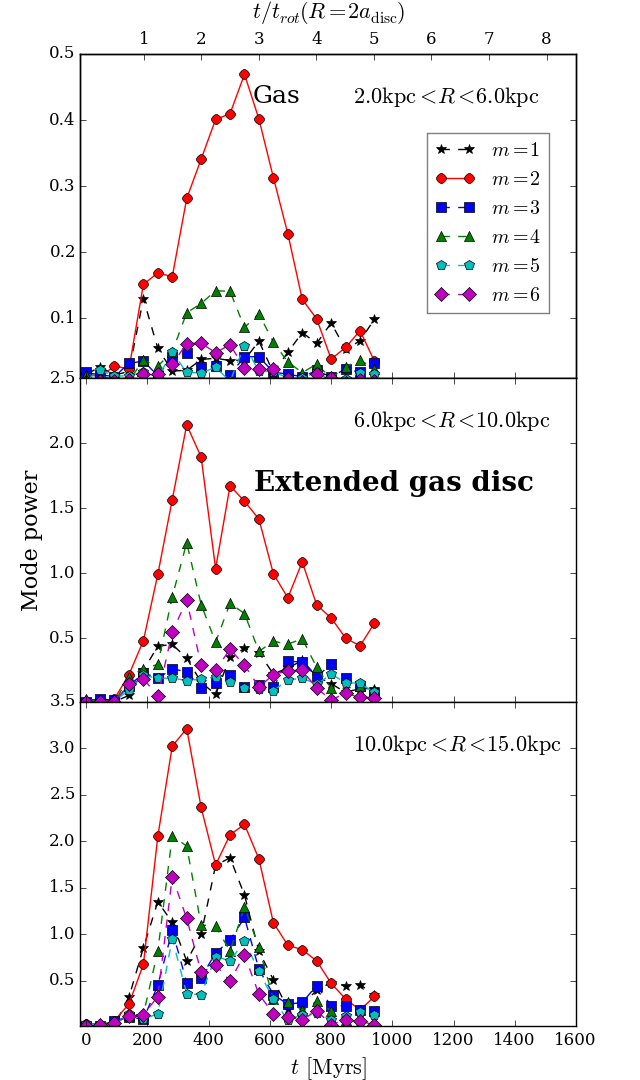}}
\resizebox{.22\hsize}{!}{\includegraphics[trim = 0mm 0mm 0mm 0mm]{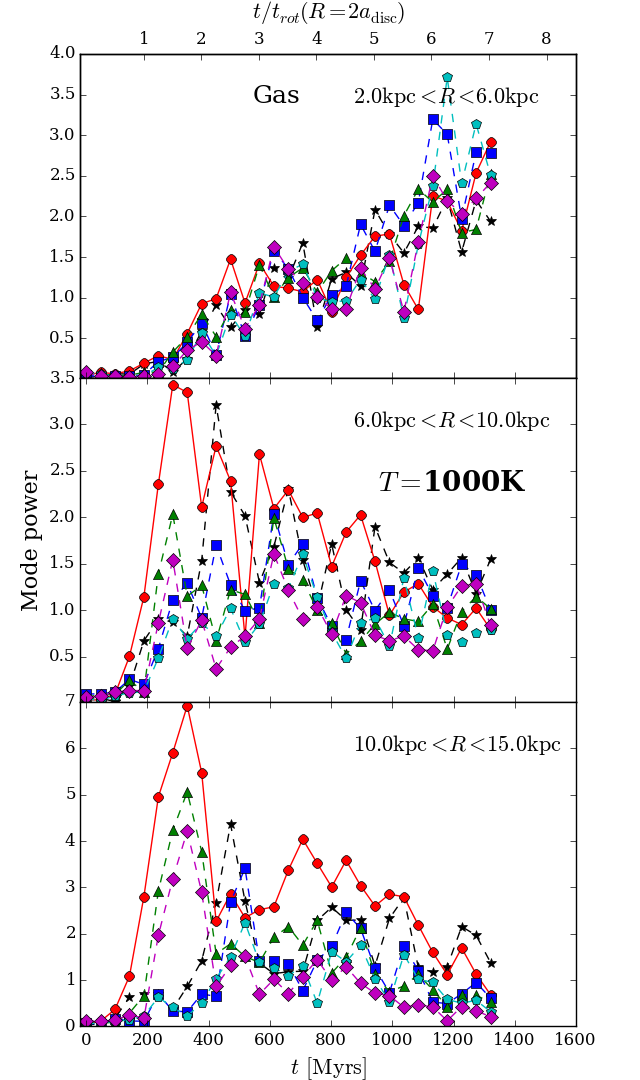}}
  \caption{As Figure 4 in the main manuscript, here we show the original run (far left, 3e5 disc star particles), the same for the higher resolution run (second column, 1e6 disc stars), for an extended gas disc (third column, $a_g=2a_d$, 3e5 disc star particles) and for the lower temperature run ($T=1000$K, far right). The power spectrum for the stars is very similar in each case, and the gas disc is similar in all but the low temperature calculation.}
  \label{ArmModesEx}
\end{figure*}

In Fig.\,\ref{Models} we show a comparison of the arms driven in three simulations with varying resolution. In the first panel is our default calculation ($3\times 10^5$ particles in the gas, stars and dark matter), second is a higher resolution run ($1\times 10^6$ particles in the gas, stars and dark matter) and third is our run with a resolved companion, using $1\times 10^4$ star particles rather than a point mass. For the 3 million particle run we use a softening of 10pc (as opposed to 50pc for the rest of our calculations). We find the arm features to be extremely similar in the three runs. The higher resolution galaxy has higher peak arm densities, and slightly clearer spurring. There is nearly no difference between the resolved or point mass companion, indicating that for small mass companions on grazing orbits, a point mass companion is an acceptable approximation.

Fig.\,\ref{ArmModesEx} shows the arm response in the stars and gas for the higher resolution model (second column), the model with an extended gas disc (third column) and the 1000K model (fourth column). The data for the fiducial calculation is shown in the first column for comparison (same as Fig.\,\ref{ArmModes}). There is only marginal difference between the response both gas and stars in the different resolution cases. For the extended gas disc, additional dynamical drag caused the companion to be strongly bound shortly after passage, and so no data is shown after 1Gyr (the time when companion re-enters the system). The main difference in the extended gas distribution is that the response in the outer disc appears weaker (bottom panel of the furthest right plot), though the general response is similar the magnitude is weaker. This is likely because the surface density of the gas in this $10{\rm kpc} < R <15{\rm kpc}$ region has been reduced by extending the gas disc out to 40kpc as opposed to the usual 20kpc, with a lower enclosed mass at the radius of closest approach reducing the tidal response in the disc. While stars in the $T=1000$K calculation display the same features as the fiducial run, the gas disc is quite different, especially in the inner disc. The initial increase in power occurs at a similar rate as the other calculations, but then continues to increase to high levels after the companion leaves the system. This is due to the higher surface density in the inner disc causing gas collapse into large cloud-like structures. These dense blobs only increase in size after their initial triggering due to the companion passage. These blobs cause rapid changes in the amplitude of different Fourier modes over small radial distances, causing the seemingly constant increase in all modes in the inner gas disc of the colder simulation.

\bsp
\label{lastpage}
\end{document}